\documentclass[sigconf,natbib=true]{acmart}
\AtBeginDocument{%
  \providecommand\BibTeX{{%
    \normalfont B\kern-0.5em{\scshape i\kern-0.25em b}\kern-0.8em\TeX}}}

\copyrightyear{2022}
\acmYear{2022}
\setcopyright{acmcopyright} \acmConference[SIGIR '22]{Proceedings of the 45th International ACM SIGIR Conference on Research and Development in Information Retrieval}{July 11--15, 2022}{Madrid, Spain}
\acmBooktitle{Proceedings of the 45th International ACM SIGIR Conference on Research and Development in Information Retrieval (SIGIR '22), July 11--15, 2022, Madrid, Spain}
\acmPrice{15.00}
\acmDOI{10.1145/3477495.3531965}
\acmISBN{978-1-4503-8732-3/22/07}

\usepackage{amsmath}
\usepackage{setspace}
\usepackage{graphicx} 
\usepackage{float} 
\usepackage{subfigure} 
\usepackage{multirow}
\usepackage{diagbox}
\usepackage{color}
\usepackage{enumitem}
\usepackage{ulem}
\usepackage{xcolor}
\begin{document}
% \renewcommand\cwnote[1]{}
%%
%% The "title" command has an optional parameter,
%% allowing the author to define a "short title" to be used in page headers.
\title{Determinantal Point Process Likelihoods \\ for Sequential Recommendation}

%%
%% The "author" command and its associated commands are used to define
%% the authors and their affiliations.
%% Of note is the shared affiliation of the first two authors, and the
%% "authornote" and "authornotemark" commands
%% used to denote shared contribution to the research.

\author{Yuli Liu}
\affiliation{%
  \institution{Australian National University \& Data61, CSIRO}
  \city{Canberra}
  \country{Australia}}
\email{yuli.liu@anu.edu.au}

\author{Christian Walder}
\affiliation{%
  \institution{Data61, CSIRO \& \\ Australian National University}
  \city{Canberra}
  \country{Australia}}
\email{christian.walder@data61.csiro.au}

\author{Lexing Xie}
\affiliation{%
  \institution{Australian National University \& Data61, CSIRO}
  \city{Canberra}
  \country{Australia}}
\email{lexing.xie@anu.edu.au}

%%
%% By default, the full list of authors will be used in the page
%% headers. Often, this list is too long, and will overlap
%% other information printed in the page headers. This command allows
%% the author to define a more concise list
%% of authors' names for this purpose.
\renewcommand{\shortauthors}{}

%%
%% The abstract is a short summary of the work to be presented in the
%% article.
\begin{abstract}
Sequential recommendation is a popular task in academic research and close to real-world application scenarios, where the goal is to predict the next action(s) of the user based on his/her previous sequence of actions. In the training process of recommender systems, the loss function plays an essential role in guiding the optimization of recommendation models to generate accurate suggestions for users. However, most existing sequential recommendation techniques focus on designing algorithms or neural network architectures, and few efforts have been made to tailor loss functions that fit naturally into the practical application scenario of sequential recommender systems. 

Ranking-based losses, such as cross-entropy and Bayesian Personalized Ranking (BPR) are widely used in the sequential recommendation area. We argue that such objective functions suffer from two inherent drawbacks: \textit{i)} the dependencies among elements of a sequence are overlooked in these loss formulations; \textit{ii)} instead of balancing accuracy (quality) and diversity, only generating accurate results has been over emphasized. We therefore propose two new loss functions based on the Determinantal Point Process (DPP) likelihood, that can be adaptively applied to estimate the subsequent item or items. The DPP-distributed item set captures natural  dependencies among temporal actions, and a quality \textit{vs.} diversity decomposition of the DPP kernel pushes us to go beyond accuracy-oriented loss functions. Experimental results using the proposed loss functions on three real-world datasets show marked improvements over state-of-the-art sequential recommendation methods in both quality and diversity metrics.

\end{abstract}

%%
%% The code below is generated by the tool at http://dl.acm.org/ccs.cfm.
%% Please copy and paste the code instead of the example below.
%%
\begin{CCSXML}
<ccs2012>
<concept>
<concept_id>10002951.10003317.10003347.10003350</concept_id>
<concept_desc>Information systems~Recommender systems</concept_desc>
<concept_significance>500</concept_significance>
</concept>
</ccs2012>
\end{CCSXML}

\ccsdesc[500]{Information systems~Recommender systems}

%%
%% Keywords. The author(s) should pick words that accurately describe
%% the work being presented. Separate the keywords with commas.
\keywords{Sequential Recommendation, Determinantal Point Process, Loss Function, Diversity}

%% A "teaser" image appears between the author and affiliation
%% information and the body of the document, and typically spans the
%% page.

%%
%% This command processes the author and affiliation and title
%% information and builds the first part of the formatted document.
\maketitle
\pagestyle{plain}

\section{INTRODUCTION}
Traditional recommender systems, \textit{e.g.}, \textit{Top-N} recommendation \cite{hu2008collaborative, deshpande2004item}, strive to model users' preferences towards items based on historical interactions between users and items. The modeling process only depends on static interactions and ignores sequential dependencies with the assumption that all user-item interactions are equally important. However, this might not hold in real-world scenarios \cite{fang2020deep}, where the next action(s) of a user are generated according to his/her previous behavioral sequence. To model users' evolving interests and adjust recommendation methods to real applications, the customized time-relevant recommender system (\textit{i.e.}, sequential recommendation) is proposed and has become a popular topic in academic research and online applications.

Sequential dependencies among items are relevant for next item(s) prediction. For example, if it is known that a user has browsed "\textit{shirt}, \textit{hat} and \textit{watch}" (as shown in Figure 1), the conventional recommender typically provides content-similar or category-similar items \cite{liang2021recommending}, such as more watches or clothes. As for sequential recommendation, considering dependencies and the temporal order of a previous behavioral sequence, the recommendation model tends to capture the temporal dependency of previous actions on target items (\textit{i.e.}, \textsl{sequence dependence}), and thereby determines that the user is actually trying to find clothing accessories (\textit{e.g.}, \textit{headphones} and \textit{glasses}) in this period of time.

The core consideration of recommending the item or items is the user preference-based ranking of items. Similar to most learning-to-rank tasks (\textit{e.g.}, Web search and question answering), commonly used ranking-based loss functions such as pointwise ranking (cross-entropy) and pairwise ranking (BPR) are widely deployed for the training of recommendation models. In this paper, we argue the insufficiency of applying ranking-based loss functions in sequential recommendation from three aspects.  

Firstly, we regard estimating the next $T$ items (\textit{i.e.}, target items in Figure 1) as a task of predicting the temporally subsequent set, according to the historical sequence. In this scenario, the dependency among targets is noteworthy. However, the ranking-based objective function independently compares each target to its true label (pointwise ranking), to a negative item (pairwise ranking), or to a perfect ordering list (listwise ranking). This manipulation neglects the dependency among targets (\textit{targets dependency}), thereby affecting the set prediction capacity of sequential recommendations. For example, based on the previous sequence in Figure 1, the loss calculated assuming independent targets (\textit{headphones} or \textit{glasses}) merely makes sequential models capture the dependencies in two sequences \{\textit{shirt}, \textit{hat}, \textit{watch}, \textit{headphones}\} or \{ \textit{shirt}, \textit{hat}, \textit{watch}, \textit{glasses}\}. This means that part of the \textsl{sequence dependency} is neglected, due to the omission of \textit{targets dependency}. On the other hand, the inherent goal of next-item recommender systems is to select or generate personalized item sets for specific users. However, ranking-based loss functions enforce the rank or estimated score of targets, leading sequential recommendations to provide high-scored items (\textit{i.e.}, \textit{top-N} prediction), without considering the feasibility of selecting whole targets as a generated set. In this work, instead of stressing the ranking of target items in loss functions, we propose to focus on enhancing the likelihood of target sets, \textit{i.e.}, endowing the target set with a high probability of being selected/recommended.

\begin{figure}
  \centering
  \includegraphics[width=0.99\linewidth]{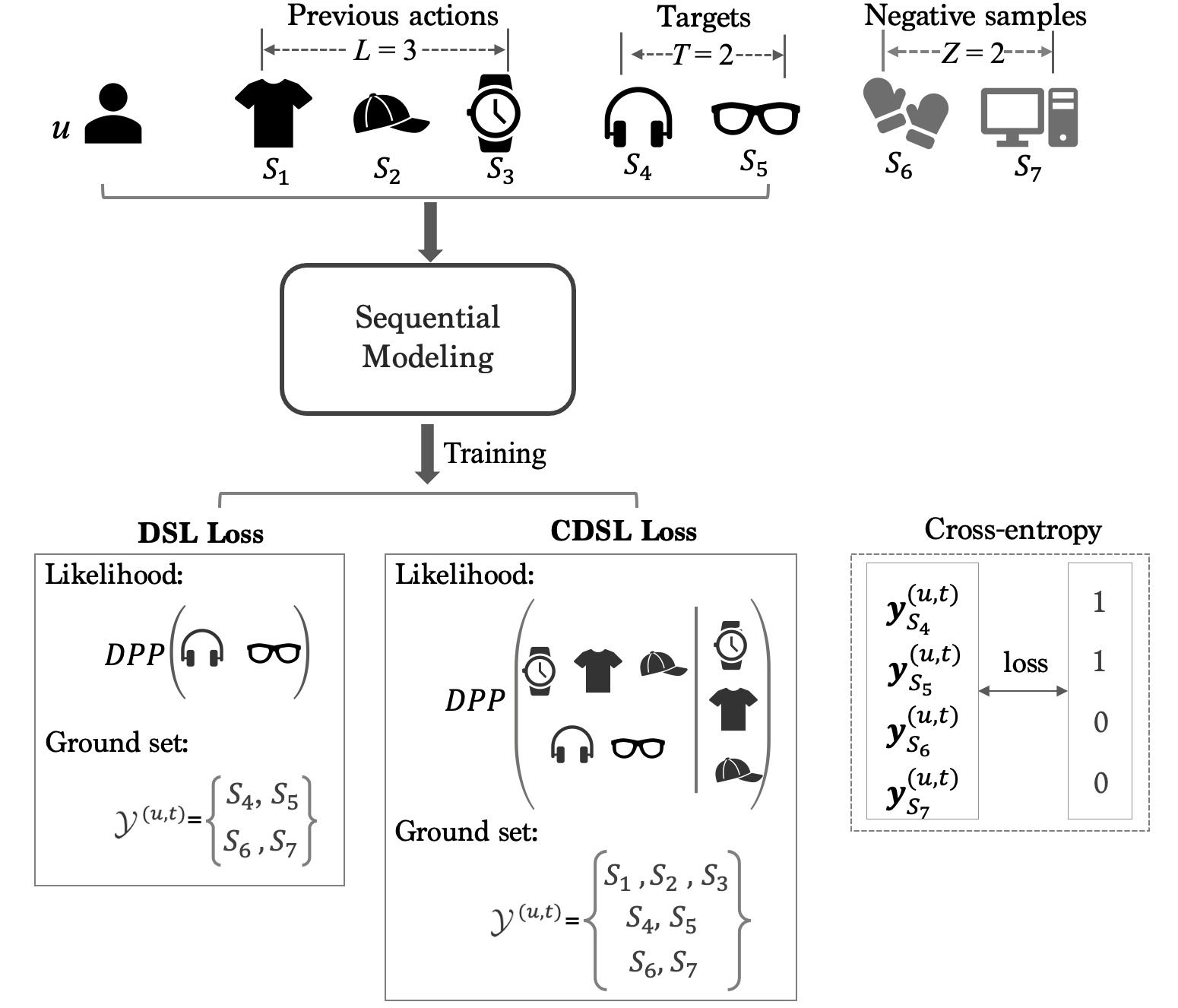}
  \caption{A simplified diagram comparing the loss functions based on DPP-distributed set likelihoods with a basic cross entropy loss function.}
  \Description{.}
  \vspace{-3mm}
\end{figure}

Based on the Determinantal Point Process (DPP) \cite{kulesza2012determinantal, kulesza2011k}, we propose the DPP Set Likelihood-based loss function (DSL). We regard the temporal set prediction task as drawing a subset from a user-sequence specified ground set (denoted as $\mathcal{Y}^{(u, t)}$ in Figure 1) according to the DPP probability measure. $\mathcal{Y}^{(u, t)}$ is the combination of target items and negative samples for each user's sequence at time step $t$. It means that the drawn subset represents the probability of inclusion for user-preferred targets and exclusion for non-relevant items. In this way, the set likelihood-based loss function considers the correlations among observed targets, and enforces that the chance of suggesting the target set should be higher than that of selecting the negative one. 

Secondly, existing sequential recommendations that try to capture the \textsl{sequence dependency} (on the entire sequence) \textit{e.g.}, \{\textit{shirt}, \textit{hat}, \textit{watch}, \textit{headphones}, \textit{glasses}\} in Figure 1, mainly focus on modeling the dependency in neural layers \cite{tang2018personalized, kang2018self} or Markov chains \cite{rendle2010factorizing, he2016fusing}. However, in a more challenging part, \textit{i.e.}, the training loss, the dependencies among temporal actions are neglected. DSL considers \textsl{targets dependency} modeling in loss calculation, but the dependency between previous actions and targets is not directly modeled. Specifically, DSL is proposed for the scenario of predicting the next items (\textit{i.e.}, $T$ > 1 in Figure 1).

To model the entire dependency of a sequence and generalize DPP likelihood-based loss to next item(s) estimation (\textit{i.e.}, predicting next one or more targets in training), a new Conditional DPP Set Likelihood (CDSL) objective is designed. We condition a DPP on the event that elements (items) in the previous set are observed. Specifically, given the previous actions, the target item(s) subset is selected (distributed as a DPP) and negative samples are excluded. In this case, the user-sequence specified ground set contains items of the entire sequence ($L+T$) and corresponding negative samples $Z$. Conditioning a DPP on observed previous items helps recommendation models capture more correlations is more powerful for subset drawing. Experiments on different datasets demonstrated these intuitions. 

Thirdly, ranking-based losses mainly aim at improving the relevance score of target items, whereas they account little for the diversity that is an important factor for avoiding users being bored with the narrow scope of topics of recommendations. This causes recommender systems to unduly lean on accuracy recommendations \cite{wu2019pd} while ignoring the diversity.

The Determinantal Point Process (DPP), as an elegant probabilistic model, is endowed with the property of modeling the repulsive correlations (dissimilarity) of elements in a subset. This property makes DPP a natural choice for diversity-promoting recommender systems, as it facilitates providing items that are diversified. Unlike previous studies that apply DPP for diversification in the process of building model architectures \cite{gartrell2019learning, warlop2019tensorized, wilhelm2018practical} or generating diverse and relevant items by maximum a posteriori (MAP) inference \cite{wu2019pd, chen2018fast}, we embed the recommendation diversity in the set likelihood-based loss functions by employing the quality (accuracy or relevance) \textit{vs.} diversity decomposition of DPP kernel. 

We summarize the contributions of this work as follows: 
\begin{itemize}
\item A new perspective (\textit{i.e.}, set selection/prediction instead of ranking items) on loss functions for sequential recommendations is contributed.
\item Two objective functions, DSL and CDSL bring \textsl{targets dependency} and \textsl{sequence dependency} into training loss and fit naturally into the next items and item(s) prediction. 
\item The diverse kernel $\mathbf{K}$ learned from the observed item sets and personalized relevance scores predicted by the sequential model play roles of diversity and quality in the diversity-aware loss functions, respectively.  
\end{itemize}

%% targets-depencency and  
%% time-relevant, sequential patterns
%% can be reworked 
%% selecting set instead ranking
%% CDSL more viable for
%%attracted substantial attention as an elegant probabilistic model that captures the balance between quality and diversity within sets.
%%critical building blocks of recommendation list optimization

\section{PROPOSED LOSS FUNCTIONS}
The simplified illustration of DPP set likelihood-based loss functions (DSL and CDSL) is shown in Figure 1. The most apparent difference between the commonly used pointwise ranking loss (binary cross entropy) and the proposed loss functions (DSL and CDSL) is that the cross entropy is a type of accuracy comparison loss \cite{ruby2020binary}, \textit{i.e.}, measuring the probability error between estimated scores and true labels, whereas DSL and CDSL calculate the loss measure by DPP set distribution comparison, \textit{i.e.}, the comparison between inclusion probability for targets and exclusion probability for negative samples. In addition, the proposed DSL and CDSL  directly apply the output results of sequential models to calculate the proposed losses and no specific changes to the model framework are required. This indicates that DSL and CDSL can be flexibly employed in different sequential models.

\subsection{Preliminaries}
Before we present the proposed loss functions, we briefly review the preliminaries of ranking-based losses and DPP.

\subsubsection{Ranking-based losses}
Like other learning to rank tasks, the core of recommender systems is the relevance-based ranking of items\cite{hidasi2015session}. Ranking can be pointwise, pairwise, or listwise. 

There are different types of pointwise-based losses, such as cross-entropy, squared loss, and MRR optimization \cite{steck2015gaussian}. In this section, we only present the binary cross-entropy loss, as different pointwise-based losses share the similar concept and cross-entropy is the most common and representative one \cite{he2017neural}.  

The binary cross-entropy loss for next items ($T$ targets) prediction is defined as
\begin{equation}
\begin{aligned}
    \ell= \sum_{u} \sum_{t \in C^{u}} \sum_{i \in \mathcal{D}_{t}^{u}}-\log \left(\sigma\left(\boldsymbol{r}_{i}^{(u, t)}\right)\right)+\sum_{j \notin \mathcal{D}_{t}^{u}}-\log \left(1-\sigma\left(\boldsymbol{r}_{j}^{(u, t)}\right)\right),
\end{aligned}
\end{equation}
where $\boldsymbol{r}_{i}^{(u, t)}$ is the estimated preference score of user $u$ on target item $i$ in the sequence of time step $t$, and $\sigma$ usually represents the $sigmoid$ activation function. $C^{u}$ is a collection of specific time steps of user $u$. To make it applicable to next items prediction, $\mathcal{D}_{t}^{u}$ is used to denote the next $T$ target items. 
The cross-entropy format can be derived from the likelihood function
\begin{equation}
\begin{aligned}
    p(\mathcal{S} \mid \Theta)=\prod_{u} \prod_{t \in \mathcal{C}^{u}} \prod_{i \in \mathcal{D}_{t}^{u}} \sigma\left(\boldsymbol{r}_{i}^{(u, t)}\right) \prod_{j \notin \mathcal{D}_{t}^{u}}\left(1-\sigma\left(\boldsymbol{r}_{j}^{(u, t)}\right)\right) 
\end{aligned}
\end{equation}
by taking the negative logarithm. From this equation we see that the cross-entropy loss independently estimates the ranking or score, ignoring dependency among target items. 

Pairwise ranking (BPR) \cite{rendle2012bpr} is also widely used in learning-to-rank tasks, formulated as:
\begin{equation}
\begin{aligned}
    \ell= \sum_{u} \sum_{t \in C^{u}} \sum_{i \in \mathcal{D}_{t}^{u}}-\ln \sigma\left(\boldsymbol{r}_{(u,i,j)}^{t}\right),
\end{aligned}
\end{equation}
where $\boldsymbol{r}_{(u,i,j)}^{t}=\boldsymbol{r}_{(u,i)}^{t}-\boldsymbol{r}_{(u,j)}^{t}$. Item $j$ is an item that user $u$ shows no interest.
This formulation is derived from the maximum a posteriori (MAP) estimator of the model parameters ($\Theta$):
\begin{equation}
\begin{aligned}
\underset{\Theta}{\operatorname{argmax}} \prod_{u} \prod_{t \in C^{u}} \prod_{i \in \mathcal{D}_{t}^{u}} p\left(i >_{u, t} j \mid \Theta\right) p(\Theta),
\end{aligned}
\end{equation}
based on the Bayesian formulation and the independence assumption of targets and users. This means that pairwise ranking loss also takes no account of targets dependency and sequence dependency. The above formulations are defined in the next $T$ targets scenario, which can be easily transformed for the next item task.

Instead of treating either each rating (cross-entropy) or each pairwise comparison (BPR) as an independent instance, listwise-based ranking \cite{cao2007learning} directly assigns a loss to an entire item list (or a \textit{Top-N} list) by employing the cross-entropy to measure the distance between a predicted list and perfect ordering. The listwise learns a ranking dependent on the position of items in the relevance score-ordered lists. However, as interactions in recommender systems are usually of binary relevance (with no perfect ordering list) and the permutation probabilities are computationally expensive, listwise ranking is not used often \cite{hidasi2015session} in sequential recommendations. Therefore, listwise ranking is not compared with here. 

\subsubsection{Determinantal Point Process}
DPP is known and widely used within the machine learning community, \textit{e.g.}, diversity-promoting recommendations \cite{wu2019pd} and neural conversation \cite{song2018towards}. It provides tractable and efficient means to balance quality and diversity within a subset. Formally, a determinantal point process $\mathcal{P}$ on a discrete set $\mathcal{Y}=\{1,2,...,M\}$ is a probability measure on $2^{\mathcal{Y}}$, the set of all subsets of $\mathcal{Y}$). When $\mathcal{P}$ gives nonzero probability to the empty set, there exists a matrix $\mathbf{L} \in \mathbb{R}^{M \times M}$, such that for every subset $Y \subseteq \mathcal{Y}$, the probability is $\mathcal{P}(Y) \propto \operatorname{det}\left(\mathbf{L}_{Y}\right)$, where $\mathbf{L}$ is a real, positive semi-definite kernel matrix indexed by the elements of $\mathcal{Y}$. In the context of sequential recommendation, $\mathcal{Y}$ is the entire item (movies or products) catalog, and $Y$ is the subset of items that users interact with. The notation $\mathbf{L}_Y$ denotes the submatrix of kernel $\mathbf{L}$ restricted to only those rows and columns which are indexed by $Y$.

The marginal probability of including one element $S_i$ is $\mathcal{P}(S_i) = \mathbf{L}_{ii} $. That is, the diagonal of $\mathbf{L}$ gives the marginal probabilities of inclusion for individual elements of $\mathcal{Y}$. The probability of selecting two items $S_{i}$ and $S_{j}$ is $\mathbf{L}_{ii}\mathbf{L}_{jj} - \mathbf{L}^{2}_
{ij} = \mathcal{P}(S_{i}) \mathcal{P}(S_{j})-\mathbf{L}_{i j}^{2}$. The entries of the kernel $\mathbf{L}$ usually measures similarity between pairs of elements in $\mathcal{Y}$. Thus, highly similar elements are unlikely to appear together. In this way, repulsive correlations (\textit{i.e.}, diversity) with respect to a similarity measure are captured. However, an important factor --- personalized relevance to items in recommendations are not reflected in the similarity-based kernel matrix. To adapt DPP to user preferences- and diversity-aware sequential recommendation tasks, we bring a quality \textit{vs.} diversity decomposition of the DPP kernel by independently modeling diversity and quality.

\subsection{Quality \textit{vs.} Diversity Kernel}
Similarity measures reflect the primary qualitative characterization of the DPP as diversifying processes. However, in practical recommendations, diversity needs to be balanced against underlying user preferences (quality) for different items. The decomposition of the DPP kernel $\mathbf{L}$ offers an intuitive trade-off between quality of items in $\mathcal{Y}$ and a global model of diversity, whose entries can be written $\mathbf{L}_{ij}=q_{i} \phi_{i}^{\top} \phi_{j} q_{j}$. $q_{i}\in \mathbb{R}^+$ is a quality term of item $i$ and $\phi_{i}\in \mathbb{R}^D$ is a D-dimensional vector of normalized diversity features (in practice, it is usually represented by feature vectors) \cite{kulesza2012determinantal}. 
In sequential recommendations, user preferences on items of a sequence can be naturally treated as the quality measure. On the other hand, we need a diversity kernel to capture the diversity of items, which can be learned from the observed diverse item sets \cite{gartrell2017low}. To reduce the computational complexity of calculating a $\mid\mathcal{Y}\mid \times \mid\mathcal{Y}\mid$ matrix, the diversity kernel is represented using the low-rank factorization as $\mathbf{K}=\mathbf{V}^{\top}\mathbf{V}$, where $\mathbf{V}$ is a $M\times D$ matrix and $D<<\mid\mathcal{Y}\mid$. From now on we use $\mathbf{K}$ to denote the \textit{diversity kernel}, and $\mathbf{L}$ is the final kernel that is balanced by quality and diversity. 

We first learn \textit{diversity kernel} $\mathbf{K}$ from observed interactions (instances in training datasets) before training sequential models. To capture the co-occurrence of diverse items, we use the following diverse sets generation process to diversify item sets:
\textit{i)} before sequentially selecting items for a diverse set, we assign each item in a user's interaction list with the same probability of being selected; 
\textit{iii)} every time an item $i$ is selected. Its category $c_i$ will be stored. The probability of all items in $c_i$ will decay ($decay=0.5$). Because items are randomly selected according to their corresponding probabilities, the decay setting facilitates generating item sets with different categories (\textit{i.e.}, diversity). 
Five distinct items are selected for each item subset;
\textit{iii)} repeat the last step until the observable positive items of the user have all been selected at least once.

For each positive item set, a corresponding negative set is  sampled, which contains negative items that share the same categories with items of the positive set, but have no interaction with the user.

Based on these ground-truth diverse sets, the diversity kernel $\mathbf{K}$ can be learned following \cite{warlop2019tensorized}, with the log-likelihood formulation:
\begin{equation}
\begin{aligned}
    \ell=\sum_{\big(T^{(+)},T^{(-)}\big)\in\mathcal T} \log \operatorname{det}\left(\mathbf{K}_{T^{(+)}}\right)-\log \operatorname{det}\left(\mathbf{K}_{T^{(-)}}\right),
\end{aligned}
\end{equation}
where $T^{(+)}$ is an observed diverse set and $T^{(-)}$ represents the corresponding set that contains negative items and $\mathcal T$ denotes the set of paired sets used for training.

In sequential recommendation, the training loss is calculated by accumulating the difference comparison between targets and predictions from each sequence of users in a batch. To compare the likelihood between the true set and negative set in each sequence, we define the following user-sequence specified  kernel:
\begin{equation}
\begin{aligned}
    \mathbf{L}^{(u, t)}=\operatorname{Diag}\left(\mathbf{R}^{(u, t)}\right) \cdot \mathbf{K}^{(u, t)} \cdot \operatorname{Diag}\left(\mathbf{R}^{(u, t)}\right),
\end{aligned}
\end{equation}
where $\mathbf{R}^{(u, t)}$ represents user $u$'s preferences on specific items (\textit{sequence ground set}, \textit{i.e.}, $\mathcal{Y}^{(u, t)}$ in Figure 1) depending on the observed sequence of time step $t$, and $\mathbf{K}^{(u, t)}$ is the submatrix of diversity kernel indexed by the \textit{sequence ground set}. Similarly, $\mathbf{L}^{(u, t)}$ is used to denote the submatrix of $\mathbf{L}$ DPP kernel. We name it as \textit{sequence kernel}, as it is confined to a user sequence. As a shorthand, we will remove the superscript $(u,t)$ that represents a specific ground set associated to a user sequence when the meaning is clear. 

\subsection{DPP Set Likelihood-Based Loss}
We first introduce the intuitive idea of using DPP set likelihood comparison as a loss function for sequential recommendations. Given a sequence, the targets and negative items can be regarded as the user-sequence specified ground set, and we can treat the measurement of loss as the negative log probability of selecting targets as a DPP. This means that the target set is supposed to receive a higher probability of being drawn by $\mathcal{P}$ (mentioned in Section 2.1.2) than that of negative items. The normalization constant for $\mathcal{P}$ follows from the observation that $\sum_{Y^{\prime} \subseteq \mathcal{Y}} \operatorname{det}\left(\mathbf{L}_{Y^{\prime}}\right)=\operatorname{det}(\mathbf{L}+\mathbf{I})$, where $\mathbf{I}$ is the identity matrix \cite{gartrell2017low}. The probability of $\operatorname{det}\left(\mathbf{L}_{Y}\right)$ is normalized by all possible item sets $Y^{\prime} \subseteq \mathcal{Y}$. Therefore, we have the following probability of inclusion for target set $Y_T$ in a user sequence:
\begin{equation}
\begin{aligned}
    \mathcal{P}(Y_{T})=\frac{\operatorname{det}\left(\mathbf{L}_{Y_{T}}\right)}{\operatorname{det}(\mathbf{L}+\mathbf{I})},
\end{aligned}
\end{equation}
The ground set of this sequence is $\mathcal{Y}^{(u,t)}$ as shown in the DSL part of Figure 1. $Y_T$ denotes target items set in a sequence. More precisely, $\mathbf{I}$ is identity matrix of the \textit{sequence ground set}, \textit{i.e.}, with ones in the diagonal positions. It represents a $\mid \mathcal{Y}^{(u,t)}\mid \times \mid \mathcal{Y}^{(u,t)}\mid$ matrix. In DSL, \textit{sequence ground set} contains targets and negative samples in a user's sequence instance. 

Based on the introduction above, we can see that DSL is expected to capture the dependency among targets. In this work, DSL is only applied to the next items (target items $T>1$) estimation. This is to the fact that if only one item is in the target set, the significance of considering \textit{targets dependency} is lost. Besides, DSL ignores the \textit{sequence dependency} (correlations between previous items and targets). To make the DPP-based loss function more viable, we further propose CDSL. 

\subsection{Conditional DPP Set Likelihood-Based Loss}
To capture entire \textit{sequence dependency} in loss functions, we try to formulate the motivation of drawing a DPP set ($T$ targets) conditioning on the observed previous set ($L$ items). Given the previous items ($Y_L$) in a user-specified sequence, we want to obtain the probability of drawing an entire sequence ($Y_{L\cup T}$), \textit{i.e.}, $L$ previous items and $T$ targets, which is formulated as:

\begin{equation}
\begin{aligned}
    \mathcal{P}(Y_{L\cup T} \mid {Y}_L) =\frac{\operatorname{det}\left(\mathbf{L}_{Y_{L\cup T}}\right)}{\operatorname{det}\left(\mathbf{L}+\mathbf{I}_{\overline{Y}_{L}}\right)},
\end{aligned}
\end{equation}
where $\mathbf{L}$ is actually the \textit{sequence kernel} on $\mathcal{Y}^{(u,t)}$ that is composed of previous items, targets and negative samples (as shown in the CDSL part of Figure 1). $\mathbf{I}_{\overline{Y}_{L}}$ is the matrix with ones in the diagonal entries indexed by elements of $\mathcal{Y}^{(u,t)}-\overline{Y}_{L}$ and zeros everywhere else. This conditional distribution is derived according to \cite{borodin2005eynard, kulesza2012determinantal}. If you are interested, you can learn more from the literature. From this equation, we can see that even if $T$ equals 1 (next item scenario), CDSL can still be applied to utilize the dependency by calculating DPP set ($Y_{L\cup T}$) probability. This means that CDSL is more adaptable.      

Based on Equation (7) and Equation (8), DSL and CDSL loss functions for next $T$ targets and next target(s) estimations can be obtained by taking the negative log-likelihood. 

\section{EXPERIMENTS}
We comprehensively evaluate the proposed DSL and CDSL loss functions in terms of \textit{i)} recommendation quality, \textit{ii)} diversification, and \textit{iii)} training efficiency.\footnote{Our source code can be found at \url{https://github.com/l-lyl/DPPLikelihoods4SeqRec}.}

\begin{table}[tp]
\centering
  \fontsize{9.3}{11}\selectfont
  \caption{Statistics of the datasets.}
  \label{tab:statistics}
    \begin{tabular}{ccccc}
    \hline
    Dataset&\#Users&\#Items&\#Interactions&\#Categories\\
    \hline
    \textit{ML-1M}& 6.0k & 3.4k & 1.0M & 18\\
    \textit{Anime}& 73.5k & 12.2k & 1.0M & 43\\
    \textit{Beauty}& 52.0k & 57.2k & 0.4M & 213\\
    \hline
    \end{tabular}
\end{table}

\subsection{Datasets}
We evaluate our methods on three datasets from three real
world applications. \textbf{MovieLens}\footnote{\url{https://grouplens.org/datasets/movielens/1m/}} is a widely used benchmark movie rating dataset. We use the version (\textit{ML-1M}) that includes one million user ratings to 18 categories of movies. \textbf{Amazon} contains a series of datasets introduced in \cite{mcauley2015image}, comprised of a large corpora of product reviews crawled from \textit{Amazon.com}. We consider \textit{Beauty} category in this work, which is notable for its high sparsity and variability. There are 213 categories of \textit{Beauty} products. \textbf{Anime}\footnote{\url{https://www.kaggle.com/CooperUnion/anime-recommendations-database}} consists of user ratings to anime in \textit{myanimelist.net}. All items in \textit{Anime} are grouped into 43 categories. The statistics of the datasets are shown in Table 1. 
The reasons why we select these three datasets are: \textit{i)} they are common public datasets in the sequential recommendation field. In particular, \textit{Beauty} and \textit{ML-1M} are also chosen as experimental datasets in two state-of-the-art baselines \cite{tang2018personalized, kang2018self} analyzed in this work. Selecting them helps to keep the comparison as fair as possible; \textit{ii)} the diversity (category coverage of recommended items) is also considered in the proposed loss functions. To evaluate the diversity metric, we need to choose datasets that contain the category attribute of items. As the category attribute is not considered in the selected baselines, we process the datasets by closely following their settings, instead of directly requiring the processed datasets from the authors. For all datasets, we use timestamps to determine the sequence order of actions. We use the 10-core setting for experiments to ensure the quality of the dataset. 

Referring to previous work \cite{kang2018self, he2017translation, rendle2010factorizing}, the most recent $T$ actions (last $T$ items) in each user’s sequence are used for testing. The remaining actions (except the $T$ items) are split into two parts: \textit{i)} first 90\% of them as the training set and \textit{ii)} the next 10\% of actions as the validation set. To better imitate the real-world sequential recommendation scenario and evaluate DSL and CDSL more comprehensively, the test set size of each dataset is not fixed but changes with the number of targets (\textit{i.e.}, $T$ in Figure 1) in training sequences. For instance, when considering the single next target ($T = 1$) in training, we only leave the last item of each user's interactions for testing, and the remaining actions are used for training and validating. Similarly, when $T = 3$, the test set contains each user's last three items. 

\subsection{Baselines}
As mentioned in Section 2, the proposed loss functions can be flexibly deployed in various sequential recommendation methods without modifying their model architecture. Under the same experimental settings, if the performance of reworked models (replacing a baseline's original loss function with DSL or CDSL) is improved, the superiority of our loss functions is demonstrated. In this paper, three baselines are taken into consideration:
\\ \textbf{Caser} \cite{tang2018personalized}: A CNN-based method that applies convolutional operations on the embedding matrix of some most recent items, achieves the \textit{state-of-the-art} sequential recommendation performance. The \textit{binary cross-entropy loss} is used for one or more targets.
\\ \textbf{MARank} \cite{yu2019multi}: A multi-order attentive neural model to extend user’s general preference by fusing individual- and union-level temporal item-item correlations is designed for next-item recommendation. BPR loss for one target is adopted.
\\ \textbf{SASRec} \cite{kang2018self}: This is a \textit{state-of-the-art} self-attention based next-item  recommendation model, which adaptively assigns weights to previous items at each time step. The \textit{binary cross-entropy loss} is calculated for each fixed-length sequence. If the length of a user's sequence is less than the maximum length $n$, \textit{‘padding’} items will be added. This means that the targets and previous items in a fixed-length sequence are both considered when calculating loss.

The above description provides three reasons for the selection of these baselines: \textit{i)} they are recently proposed state-of-the-art sequential recommendation methods; \textit{ii)} they are representative works, and many new models are built based on them \cite{sun2019bert4rec, xu2019recurrent, li2020time, li2021lightweight}; and \textit{iii)} different settings of loss function are used. As our loss functions are designed for the sequential recommendation scenario with the intuition of capturing sequential dependencies, general recommendation models are not selected as baselines. In addition, it has been demonstrated that non-sequential models that neglect the sequence order of actions are inferior to sequential models in dealing with sequential recommendation tasks \cite{tang2018personalized, yu2019multi}. 

\vspace{-2mm}
\subsection{Performance Comparison}
\begin{table*}[tp]
\centering
  \fontsize{8.}{9.6}\selectfont
  \caption{Performance comparison based on \textbf{Caser}. Significant improvements over Caser under the same setting are designated as $^{*}$ >5\%, $^{**}$ >10\%, and $^{***}$ >20\%. The bold values, marked \uwave{\ \ \ \ } and \underline{\ \ \ \ } denote the Caser achieves better performance than DSL and CDSL, than DSL and than CDSL, respectively.}
  \vspace{-1mm}
  \setlength{\tabcolsep}{1.36mm}{
  \label{tab:performance_comparison_caser}
    \begin{tabular}{|c|c|c|llllll|lll|lll|}
    \hline
    \multicolumn{2}{|c|}{\multirow{2}{*}{Dataset}}&\multirow{2}{*}{Method}&\multicolumn{6}{c|}{Quality}&\multicolumn{3}{c|}{Diversity}&\multicolumn{3}{c|}{Trade-off}\\\cline{4-15}
    \multicolumn{2}{|c|}{} & & Re@3&Re@5&Re@10&Nd@3&Nd@5&Nd@10&CC@3&CC@5&CC@10&F@3&F@5&F@10  \cr\hline
    \multirow{9}{*}{ML-1M} &  \multirow{3}{*}{$T$=1}
    &Caser &0.0595&0.0892&\underline{0.1398}&0.0430&0.0551&0.0713&0.2493&0.3493&0.5086&0.0850&0.1196& 0.1748 \\ 
     &&DSL & \hspace{0.27cm}-&\hspace{0.27cm}-&\hspace{0.27cm}-&\hspace{0.27cm}-&\hspace{0.27cm}-&\hspace{0.27cm}-&\hspace{0.27cm}-&\hspace{0.27cm}-&\hspace{0.27cm}-&\hspace{0.27cm}-&\hspace{0.27cm}-&\hspace{0.27cm}-\\
    &&CDSL &0.0636\boldmath{$^*$}&0.0905&0.1388&0.0460\boldmath{$^*$}&0.0571&0.0724&0.2618\boldmath{$^*$}&0.3632&0.5277&0.0906\boldmath{$^*$}&0.1227&0.1760\\
    \cline{2-15}
    &  \multirow{3}{*}{$T$=3}
    &Caser &0.0506&0.0768&0.1313&0.1034&0.1294&0.1634&0.2501&\uwave{0.3498}&\uwave{0.5045}&0.1177& 0.1593& 0.2281 \\
    &&DSL &0.0551\boldmath{$^*$}&0.0825\boldmath{$^*$}&0.1369&0.1129\boldmath{$^*$}&0.1400\boldmath{$^*$}&0.1756\boldmath{$^*$}&0.2505&0.3466&0.4969&0.1258\boldmath{$^*$}& 0.1684\boldmath{$^*$}& 0.2377 \\
    &&CDSL &0.0561\boldmath{$^{**}$}&0.0783&0.1329&0.1151\boldmath{$^{**}$}&0.1362\boldmath{$^*$}&0.1718\boldmath{$^*$}&0.2581&0.3570&0.5118&0.1285\boldmath{$^*$}& 0.1649& 0.2348 \\
    \cline{2-15}
    &  \multirow{3}{*}{$T$=5} 
    &Caser &0.0432&0.0678&0.1151&0.1467&0.1805&0.2210&0.2549&\textbf{0.3640}&\textbf{0.5371}&0.1383& 0.1852& 0.2560 \\
    &&DSL &0.0458\boldmath{$^{*}$}&0.0723\boldmath{$^{*}$}&0.1204&0.1513&0.1878&0.2284&0.2533&0.3604&0.5346&0.1419& 0.1911& 0.2630 \\
    &&CDSL &0.0459\boldmath{$^{*}$}&0.0694&0.1195&0.1553\boldmath{$^{*}$}&0.1861&0.2296&0.2585&0.3616&0.5326&0.1448& 0.1888& 0.2629 \\
    \hline
     \hline
    \multirow{9}{*}{Anime} &  \multirow{3}{*}{$T$=1}
    &Caser &0.2381&0.3231&0.4443&0.1829&0.2179&0.2572&\underline{0.2548}&\underline{0.3517}&\underline{0.4947}&0.2305& 0.3058& 0.4105 \\
    &&DSL & \hspace{0.27cm}-&\hspace{0.27cm}-&\hspace{0.27cm}-&\hspace{0.27cm}-&\hspace{0.27cm}-&\hspace{0.27cm}-&\hspace{0.27cm}-&\hspace{0.27cm}-&\hspace{0.27cm}-&\hspace{0.27cm}-&\hspace{0.27cm}-&\hspace{0.27cm}- \\
    &&CDSL &0.2700\boldmath{$^{**}$}&0.3452\boldmath{$^{*}$}&0.4681\boldmath{$^{*}$}&0.2115\boldmath{$^{**}$}&0.2423\boldmath{$^{**}$}&0.2820\boldmath{$^{*}$}&0.2519&0.3400&0.4866&0.2462\boldmath{$^{*}$}& 0.3152& 0.4236 \\
    \cline{2-15} 
    &  \multirow{3}{*}{$T$=3}
    &Caser &0.1667&0.2326&0.3332&0.3553&0.3770&0.4144&0.2548&0.3463&0.4858&0.2579 & 0.3242& 0.4225 \\
    &&DSL &0.1814\boldmath{$^{*}$}&0.2567\boldmath{$^{**}$}&0.3643\boldmath{$^{*}$}&0.3588&0.4035\boldmath{$^{*}$}&0.4397$^{*}$&0.2636&0.3606&0.4971&0.2668 & 0.3447\boldmath{$^{*}$}& 0.4445\boldmath{$^{*}$} \\
    &&CDSL &0.1846\boldmath{$^{**}$}&0.2636\boldmath{$^{**}$}&0.3841\boldmath{$^{**}$}&0.3736\boldmath{$^{*}$}&0.4190\boldmath{$^{**}$}&0.4589\boldmath{$^{**}$}&0.2591&0.3539&0.4877&0.2687 & 0.3474\boldmath{$^{*}$}& 0.4522\boldmath{$^{*}$} \\
    \cline{2-15}
    &  \multirow{3}{*}{$T$=5}
    &Caser &0.1456&0.1977&0.2917&0.4226&0.4636&0.4990&\textbf{0.2581}&0.3541&0.4971&0.2705 & 0.3420& 0.4404 \\
    &&DSL &0.1521&0.2134\boldmath{$^{*}$}&0.3170\boldmath{$^{*}$}&0.4575\boldmath{$^{*}$}&0.4984\boldmath{$^{*}$}&0.5292\boldmath{$^{*}$}&0.2550&0.3546&0.4979&0.2777& 0.3552& 0.4575 \\
    &&CDSL &0.1537\boldmath{$^{*}$}&0.2140\boldmath{$^{*}$}&0.3160\boldmath{$^{*}$}&0.4601\boldmath{$^{*}$}&0.5043\boldmath{$^{*}$}&0.5320\boldmath{$^{*}$}&0.2551&0.3590&0.4926&0.2786& 0.3591\boldmath{$^{*}$}& 0.4557 \\
    \hline
    \hline
    \multirow{9}{*}{Beauty}  &  \multirow{3}{*}{$T$=1}
    &Caser &0.0846&0.1076&0.1508&0.0538&0.0632&0.0806&0.0422&0.0614&0.1012&0.0524& 0.0718& 0.1080 \\
    &&DSL  & \hspace{0.27cm}-&\hspace{0.27cm}-&\hspace{0.27cm}-&\hspace{0.27cm}-&\hspace{0.27cm}-&\hspace{0.27cm}-&\hspace{0.27cm}-&\hspace{0.27cm}-&\hspace{0.27cm}-&\hspace{0.27cm}-&\hspace{0.27cm}-&\hspace{0.27cm}- \\
    &&CDSL &0.0944\boldmath{$^{**}$}&0.1310\boldmath{$^{***}$}&0.1817\boldmath{$^{***}$}&0.0716\boldmath{$^{***}$}&0.0864\boldmath{$^{***}$}&0.1029\boldmath{$^{***}$}&0.0441&0.0639&0.1034&0.0576\boldmath{$^{**}$}& 0.0805\boldmath{$^{**}$}& 0.1198\boldmath{$^{**}$} \\
   \cline{2-15}
    &  \multirow{3}{*}{$T$=3}
    &Caser &0.0414&0.0615&0.0975&0.0782&0.0928&0.1035&\textbf{0.0463}&0.0662&0.0994&0.0528& 0.0703& 0.0999 \\
    &&DSL &0.0568\boldmath{$^{***}$}&0.0840\boldmath{$^{***}$}&0.1388\boldmath{$^{***}$}&0.1045\boldmath{$^{***}$}&0.1209\boldmath{$^{***}$}&0.1420\boldmath{$^{***}$}&0.0459&0.0668&0.1073&0.0585\boldmath{$^{**}$}& 0.0809\boldmath{$^{**}$}& 0.1216\boldmath{$^{***}$} \\
    &&CDSL &0.0496\boldmath{$^{***}$}&0.0760\boldmath{$^{***}$}&0.1289\boldmath{$^{***}$}&0.0831&0.1024\boldmath{$^{**}$}&0.1260$^{***}$&0.0452&0.0657&0.1074&0.0538& 0.0757\boldmath{$^{*}$}& 0.1166\boldmath{$^{**}$} \\
    \cline{2-15}
    &  \multirow{3}{*}{$T$=5}
    &Caser &0.0434&0.0677&0.1219&0.1258&0.1472&0.1758&0.0433&\uwave{0.0637}&\uwave{0.1032}&0.0573& 0.0800& 0.1219 \\
    &&DSL &0.0482\boldmath{$^{**}$}&0.0727\boldmath{$^{*}$}&0.1306\boldmath{$^{*}$}&0.1320&0.1568$^{*}$&0.1822&0.0441&0.0629&0.1008&0.0592& 0.0813& 0.1226 \\
    &&CDSL &0.0554\boldmath{$^{***}$}&0.0837\boldmath{$^{***}$}&0.1430\boldmath{$^{**}$}&0.1468\boldmath{$^{**}$}&0.1683\boldmath{$^{**}$}&0.1954\boldmath{$^{**}$}&0.0440&0.0640&0.1052&0.0613\boldmath{$^{*}$}& 0.0849\boldmath{$^{*}$}& 0.1297\boldmath{$^{*}$} \\
    \hline
    \end{tabular}}
    \vspace{-2mm}
\end{table*}

\begin{table*}[tp]
\centering
  \fontsize{8.}{9.6}\selectfont
  \caption{Performance comparison based on \textbf{MARank}. Significant improvements over MARank under the same setting are designated as $^{*}$ >5\%, $^{**}$ >10\%, and $^{***}$ >20\%. The bold values, marked \uwave{\ \ \ \ } and \underline{\ \ \ \ } denote the MARank achieves better performance than DSL and CDSL, than DSL and than CDSL, respectively.}
  \vspace{-1mm}
  \setlength{\tabcolsep}{1.34mm}{
  \label{tab:performance_comparison_marank}
    \begin{tabular}{|c|c|c|llllll|lll|lll|}
    \hline
    \multicolumn{2}{|c|}{\multirow{2}{*}{Dataset}}&\multirow{2}{*}{Method}&\multicolumn{6}{c|}{Quality}&\multicolumn{3}{c|}{Diversity}&\multicolumn{3}{c|}{Trade-off}\\\cline{4-15}
    \multicolumn{2}{|c|}{} & & Re@3&Re@5&Re@10&Nd@3&Nd@5&Nd@10&CC@3&CC@5&CC@10&F@3&F@5&F@10  \cr\hline
    \multirow{6}{*}{ML-1M} &  \multirow{3}{*}{$T$=1}
    &MARank &0.0582&0.0864&0.1470&0.0437&0.0552&0.0746&0.2627&0.3716&0.5407&0.0853& 0.1189& 0.1839 \\ 
    &&DSL &\hspace{0.27cm}-&\hspace{0.27cm}-&\hspace{0.27cm}-&\hspace{0.27cm}-&\hspace{0.27cm}-&\hspace{0.27cm}-&\hspace{0.27cm}-&\hspace{0.27cm}-&\hspace{0.27cm}-&\hspace{0.27cm}-&\hspace{0.27cm}-&\hspace{0.27cm}- \\
    &&CDSL &0.0679\boldmath{$^{**}$}&0.1017\boldmath{$^{**}$}&0.1622\boldmath{$^{**}$}&0.0500\boldmath{$^{**}$}&0.0639\boldmath{$^{**}$}&0.0832\boldmath{$^{**}$}&0.2711&0.3798&0.5504&0.0968\boldmath{$^{**}$}& 0.1360\boldmath{$^{**}$}& 0.2007\boldmath{$^{*}$} \\
    \cline{2-15}
    &  \multirow{3}{*}{$T$=3}
    &MARank &0.0498&0.0742&0.1222&0.1066&0.1308&0.1645&0.2651&0.3745&0.5451&0.1208& 0.1609& 0.2270 \\
    &&DSL &0.0524\boldmath{$^{*}$}&0.0779&0.1303\boldmath{$^{*}$}&0.1105&0.1352&0.1687&0.2684&0.3759&0.5456&0.1250& 0.1660& 0.2347 \\
    &&CDSL &0.0528\boldmath{$^{*}$}&0.0773\boldmath{$^{*}$}&0.1308&0.1093&0.1335&0.1701&0.2743&0.3848&0.5535&0.1251& 0.1655& 0.2366 \\
    \cline{2-15}
    \hline
     \hline
    \multirow{6}{*}{Anime} &  \multirow{3}{*}{$T$=1}
    &MARank &0.1875&0.2518&0.3737&0.1433&0.1696&0.2090&\underline{0.2664}&\underline{0.3536}&0.4696&0.2041& 0.2641& 0.3596 \\
    &&DSL &\hspace{0.27cm}-&\hspace{0.27cm}-&\hspace{0.27cm}-&\hspace{0.27cm}-&\hspace{0.27cm}-&\hspace{0.27cm}-&\hspace{0.27cm}-&\hspace{0.27cm}-&\hspace{0.27cm}-&\hspace{0.27cm}-&\hspace{0.27cm}-&\hspace{0.27cm}- \\
    &&CDSL &0.2211\boldmath{$^{**}$}&0.3039\boldmath{$^{***}$}&0.4453\boldmath{$^{**}$}&0.1683\boldmath{$^{**}$}&0.2023\boldmath{$^{**}$}&0.2477\boldmath{$^{**}$}&0.2617&0.3346&0.4734&0.2233\boldmath{$^{*}$}& 0.2882\boldmath{$^{*}$}& 0.4001\boldmath{$^{**}$} \\
    \cline{2-15} 
    &  \multirow{3}{*}{$T$=3}
    &MARank &0.1346&0.1731&0.2616&0.2542&0.2930&0.3361&\textbf{0.2521}&\textbf{0.3381}&\textbf{0.4585}&0.2195& 0.2759& 0.3618 \\
    &&DSL &0.1876\boldmath{$^{***}$}&0.2573\boldmath{$^{***}$}&0.3826\boldmath{$^{***}$}&0.3644\boldmath{$^{***}$}&0.4072\boldmath{$^{***}$}&0.4487\boldmath{$^{***}$}&0.2476&0.3327&0.4570&0.2610\boldmath{$^{**}$}& 0.3325\boldmath{$^{***}$}& 0.4353\boldmath{$^{***}$} \\
    &&CDSL &0.1738\boldmath{$^{***}$}&0.2417\boldmath{$^{***}$}&0.3591\boldmath{$^{***}$}&0.3361\boldmath{$^{***}$}&0.3805\boldmath{$^{***}$}&0.4264\boldmath{$^{***}$}&0.2458&0.3320&0.4517&0.2503\boldmath{$^{**}$}& 0.3212\boldmath{$^{**}$}& 0.4202\boldmath{$^{**}$} \\
    \cline{2-15}
    \hline
    \hline
    \multirow{6}{*}{Beauty}  &  \multirow{3}{*}{$T$=1}
    &MARank &0.1103&0.1524&\underline{0.2310}&0.0830&0.1004&0.1259&\underline{0.0423}&\underline{0.0612}&0.0967&0.0588& 0.0825& 0.1254 \\
    &&DSL &\hspace{0.27cm}-&\hspace{0.27cm}-&\hspace{0.27cm}-&\hspace{0.27cm}-&\hspace{0.27cm}-&\hspace{0.27cm}-&\hspace{0.27cm}-&\hspace{0.27cm}-&\hspace{0.27cm}-&\hspace{0.27cm}-&\hspace{0.27cm}-&\hspace{0.27cm}- \\
    &&CDSL &0.1183\boldmath{$^{*}$}&0.1595&0.2214&0.0925\boldmath{$^{**}$}&0.1092\boldmath{$^{*}$}&0.1292&0.0415&0.0603&0.0981&0.0596& 0.0832& 0.1258 \\
   \cline{2-15}
    &  \multirow{3}{*}{$T$=3}
    &MARank &0.0543&0.0830&\textbf{0.1542}&0.0966&0.1138&\textbf{0.1434}&0.0417&0.0602&0.0964&0.0537& 0.0747& \textbf{0.1170} \\
    &&DSL &0.0606\boldmath{$^{**}$}&0.0889\boldmath{$^{*}$}&0.1428&0.1001&0.1201\boldmath{$^{*}$}&0.1413&0.0421&0.0611&0.0991&0.0553& 0.0771& 0.1168 \\
    &&CDSL &0.0552&0.0856&0.1348&0.0975&0.1156&0.1324&0.0425&0.0603&0.0970&0.0546& 0.0754& 0.1124 \\
    \hline
    \end{tabular}}
    \vspace{-3mm}
\end{table*}

\begin{table*}[tp]
\centering
  \fontsize{8.}{9.6}\selectfont
  \caption{Performance comparison based on \textbf{SASRec}. Significant improvements over SASRec under the same setting are designated as $^{*}$ >5\%, $^{**}$ >10\%, and $^{***}$ >20\%. The bold values, marked \uwave{\ \ \ \ } and \underline{\ \ \ \ } denote the SASRec achieves better performance than DSL and CDSL, than DSL and than CDSL, respectively.}
  \setlength{\tabcolsep}{1.5mm}{
  \label{tab:performance_comparison_sasrec}
    \begin{tabular}{|c|c|c|llllll|lll|lll|}
    \hline
    \multicolumn{2}{|c|}{\multirow{2}{*}{Dataset}}&\multirow{2}{*}{Method}&\multicolumn{6}{c|}{Quality}&\multicolumn{3}{c|}{Diversity}&\multicolumn{3}{c|}{Trade-off}\\\cline{4-15}
    \multicolumn{2}{|c|}{} & & Re@3&Re@5&Re@10&Nd@3&Nd@5&Nd@10&CC@3&CC@5&CC@10&F@3&F@5&F@10  \cr\hline
    \multirow{6}{*}{ML-1M} &  \multirow{3}{*}{$T$=1}
    &SASRec &0.0484&0.0725&0.1208&0.0351&0.0442&0.0601&0.2526&0.3325&0.4616&0.0717& 0.0993& 0.1513 \\ 
    &&DSL & \hspace{0.27cm}-&\hspace{0.27cm}-&\hspace{0.27cm}-&\hspace{0.27cm}-&\hspace{0.27cm}-&\hspace{0.27cm}-&\hspace{0.27cm}-&\hspace{0.27cm}-&\hspace{0.27cm}-&\hspace{0.27cm}-&\hspace{0.27cm}-&\hspace{0.27cm}-\\
    &&CDSL &0.0490&0.0730&0.1217&0.0354&0.0449&0.0604&0.2732\boldmath{$^{**}$}&0.3522\boldmath{$^{**}$}&0.4755&0.0731& 0.1010& 0.1528 \\
    \cline{2-15}
    &  \multirow{3}{*}{$T$=3}
    &SASRec &0.0330&0.0501&\textbf{0.0891}&0.0596&0.0792&\underline{0.1110}&0.2656&0.3480&0.4804&0.0789& 0.1090& \underline{0.1656} \\
    &&DSL &0.0350\boldmath{$^{*}$}&0.0519&0.0888&0.0658\boldmath{$^{**}$}&0.0827&0.1131&0.2665&0.3511&0.4847&0.0848\boldmath{$^{*}$}& 0.1129& 0.1671 \\
    &&CDSL &0.0331&0.0507&0.0869&0.0613&0.0794&0.1089&0.2707&0.3543&0.4880&0.0804& 0.1098& 0.1631 \\
    \cline{2-15}
    \hline
     \hline
    \multirow{6}{*}{Anime} &  \multirow{3}{*}{$T$=1}
    &SASRec &0.2628&0.3593&0.5108&0.2007&0.2402&0.2888&0.2293&0.3093&0.4336&0.2305& 0.3045& 0.4160 \\
    &&DSL & \hspace{0.27cm}-&\hspace{0.27cm}-&\hspace{0.27cm}-&\hspace{0.27cm}-&\hspace{0.27cm}-&\hspace{0.27cm}-&\hspace{0.27cm}-&\hspace{0.27cm}-&\hspace{0.27cm}-&\hspace{0.27cm}-&\hspace{0.27cm}-&\hspace{0.27cm}-\\
    &&CDSL &0.2777\boldmath{$^{*}$}&0.3881\boldmath{$^{*}$}&0.5235&0.2160\boldmath{$^{*}$}&0.2611\boldmath{$^{*}$}&0.3049\boldmath{$^{*}$}&0.2339&0.3165&0.4459&0.2402& 0.3205\boldmath{$^{*}$}& 0.4295 \\
    \cline{2-15} 
    &  \multirow{3}{*}{$T$=3}
    &SASRec &0.1677&0.2324&0.3505&0.2510&0.2985&\textbf{0.3809}&0.2331&0.3152&0.4462&0.2205& 0.2882& \uwave{0.4019} \\
    &&DSL &0.1749&0.2334&0.3705\boldmath{$^{*}$}&0.2413&0.2855&0.3642&0.2406&0.3250&0.4419&0.2232& 0.2885& 0.4012 \\
    &&CDSL &0.1865\boldmath{$^{**}$}&0.2413&0.3530&0.2654&0.3086&0.3761&0.2384&0.3223&0.4493&0.2320\boldmath{$^{*}$}& 0.2967& 0.4025 \\
    \cline{2-15}
    \hline
    \hline
    \multirow{6}{*}{Beauty}  &  \multirow{3}{*}{$T$=1}
    &SASRec &0.0873&0.1238&0.2079&0.0665&0.0815&0.1086&0.0379&0.0527&0.0782&0.0508& 0.0696& 0.1047 \\
    &&DSL & \hspace{0.27cm}-&\hspace{0.27cm}-&\hspace{0.27cm}-&\hspace{0.27cm}-&\hspace{0.27cm}-&\hspace{0.27cm}-&\hspace{0.27cm}-&\hspace{0.27cm}-&\hspace{0.27cm}-&\hspace{0.27cm}-&\hspace{0.27cm}-&\hspace{0.27cm}-\\
    &&CDSL &0.1032\boldmath{$^{**}$}&0.1421\boldmath{$^{**}$}&0.2286&0.0804\boldmath{$^{***}$}&0.0962\boldmath{$^{**}$}&0.1239\boldmath{$^{**}$}&0.0386&0.0507&0.0780&0.0543\boldmath{$^{*}$}& 0.0711& 0.1081 \\
   \cline{2-15}
    &  \multirow{3}{*}{$T$=3}
    &SASRec &0.0645&0.0990&\uwave{0.1627}&0.0829&\uwave{0.1103}&0.1425&\textbf{0.0394}&0.0532&0.0808&0.0513& 0.0705& 0.1056 \\
    &&DSL &0.0686\boldmath{$^{*}$}&0.0992&0.1565&0.0894\boldmath{$^{*}$}&0.1101&0.1433&0.0383&0.0533&0.0828&0.0516& 0.0706& 0.1066 \\
    &&CDSL &0.0679\boldmath{$^{*}$}&0.1072\boldmath{$^{*}$}&0.1646&0.0909\boldmath{$^{**}$}&0.1187\boldmath{$^{*}$}&0.1528\boldmath{$^{*}$}&0.0389&0.0547&0.0834&0.0522& 0.0737& 0.1093 \\
    \hline
    \end{tabular}}
    \vspace{-2mm}
\end{table*}

To explore the influences of \textit{sequence dependency} and \textit{targets dependency}, as well as to analyze performances of the proposed loss functions in coping with different length of targets (next item and next items), we set different numbers of last items (\textit{i.e.}, $T$ ) of the training sequence instances as targets that receive estimated relevance scores for loss calculation. The target length $T$ is in $\{1, 2, 3, 5\}$. As mentioned before, we divide datasets according to the target length. For each real-world dataset (\textit{ML-1M}, \textit{Anime}, or \textit{Beauty}), the last 1, 2, 3, and 5 items of each user's temporal interactions are left for testing, and remaining actions for training and validation. Therefore we obtain four groups (1, 2, 3, and 5 unobserved items) of divided data (test, training, and validation) from each real-world dataset. Correspondingly, four groups of diverse item sets are observed from different training data of each dataset, which are separately used to learn the corresponding diverse kernel matrix $\mathbf{K}$. In this setting, we imitate the scenarios of next item and next $T$ items recommendation, as well as test the ability of DSL and CDSL to cope with different sizes of missing actions. 

Three groups of evaluation metrics are adopted for the comprehensive analysis of the recommendation performance: 
\\ \textbf{Quality}. We use two common \textit{Top-N} metrics, Recall@N (Re) and NDCG@N (Nd) to evaluate the quality (relevance) of top-scored results, N $\in \{3,5,10\}$. 
\\ \textbf{Diversity}. Diversification is also considered in the proposed loss functions. Therefore, we evaluate the recommendation diversity by the widely used metric --- category coverage (CC), which is calculated by the number of categories covered by \textit{top-N} items divided by the total number of categories available in the dataset \cite{wu2019pd, puthiya2016coverage}. A higher value of CC@N means the \textit{Top-N} items are more diverse. 
\\ \textbf{Trade-off}. To evaluate the overall performance in quality and diversity, a harmonic F-score metric (F) is employed \cite{cheng2017learning}, where F@N = $2 \times quality$@N$\times  diversity$@N / ($quality$@N + $diversity$@N). And $quality$@N represents the mean of Re@N and Nd@N, and $diversity$@N denotes CC@N. 

Following previous work \cite{wang2019neural, wang2018chat, liu2021contrastive}, an early stopping strategy is applied to avoid overfitting, \textit{i.e.}, for each experiment we stop training if Nd@5 on the validation set does not increase for 10 successive epochs. To demonstrate the superior effectiveness of our loss functions that aim at sequential recommendation tasks, we deploy DSL and CDSL in three representative works (baselines). Table 2 reports the comparison of experimental results between the reworked models and \textbf{Caser} under different settings of target length ($T$). DSL and CDSL in Table 2 represent the reworked models that are constructed through replacing the original \textit{binary cross-entropy} loss of Caser by DSL and CDSL, respectively. Similarly, Table 3 and 4 present performance comparisons based on \textbf{MARank} and \textbf{SASRec}, respectively. Under a setting $T$ of a dataset, experiments of the baseline and reworked models are carried out using the same training instances and divided data. 

Based on these three tables, we aim to demonstrate that deploying DSL or CDSL in existing sequential recommendation models can achieve better performance than using other loss functions, thus indicating the superiority of our loss functions. Therefore, tuning model parameters is not one of our focuses. In each comparison, we directly use the default model parameters of the baseline. This means that we keep the comparison fair. As Figure 1 shows, the number of negative samples ($Z$) is an important component of the user sequence ground set, which is also necessary for the selected baselines and most other related models \cite{he2016fusing, sun2019bert4rec, hu2019sets2sets} to calculate the loss. For all experiments of baselines and reworked models, we set $Z=2$ when $T=1$ and $Z=T$ in other cases. The effects of $Z$ on performance are also analyzed and compared based on Figure 2. Besides, the previous item length $L$ equals 5 when $T=1$, and $L=6$ when $T>1$ to ensure that there are enough training sequences. 

In Table 2-4, some important results are marked for clear comparison and emphasis. The value in bold means that a baseline achieves better performance than both DSL and CDSL on an experimental dataset under the setting of $T$ ($T$ test items and $T$ targets). Besides, the marker \uwave{\ \ \ \ } of a result denotes that a baseline only achieves the higher value than DSL, and the underline means that a baseline beats CDSL under a specific setting. The superscripts $^*$, $^{**}$, and $^{***}$ indicate that a reworked model (DSL or CDSL) improves by >5\%, >10\%, and >20\%, respectively over the corresponding baseline \textit{w.r.t.} a metric (in the same column) under a target length \textit{T}. For instance, in Table 2 when \textit{T} = 3 of \textit{ML-1M} dataset, the Re@3 result of CDSL (Caser with CDSL loss) shows 0.0561, which is 10.87\% higher than baseline Caser's Re@3 result (0.0506). Therefore 0.0561$^{**}$ is presented.

\begin{figure*}
\centering
    \subfigure[ML-1M dataset]{
    \includegraphics[width=0.45\linewidth]{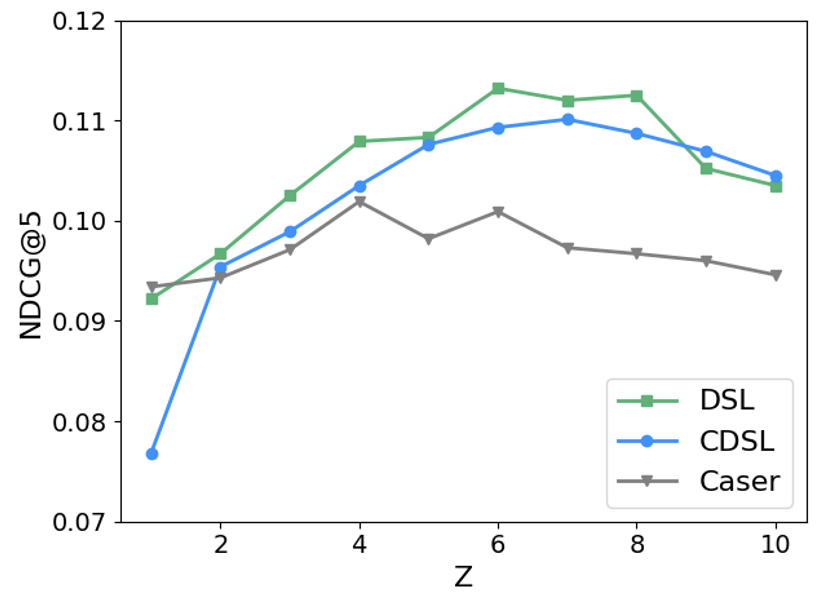}}
    \subfigure[Beauty dataset]{
    \includegraphics[width=0.45\linewidth]{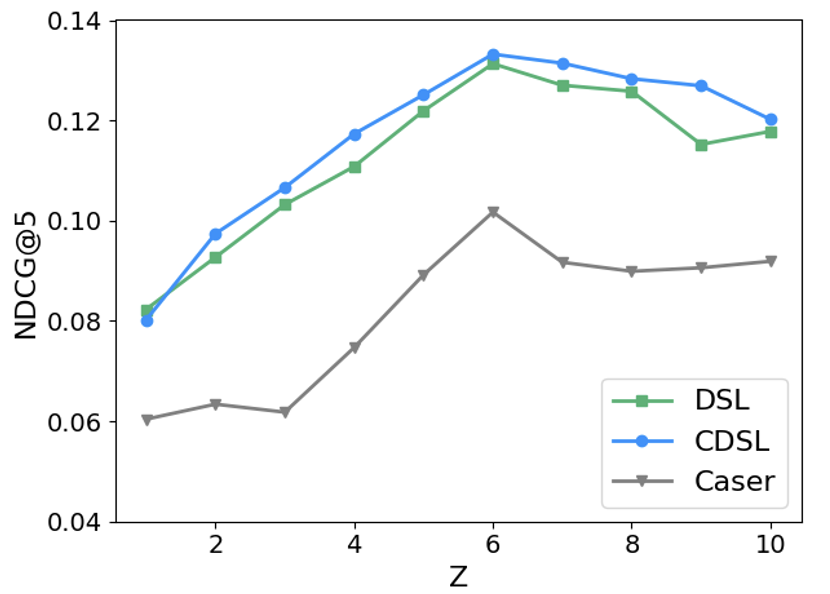}}
\vspace{-1mm}
\caption{Affects of negative sample length $Z$ on ML-1M and Beauty datasets.}
\vspace{-3mm}
\end{figure*}

We can see that there is no DSL result when \textit{T} = 1 in Table 2-4. This is because DSL is designed for the next $T$ targets estimation as explained in Section 2.3. When there is only one target in loss calculation, DSL cannot capture the targets dependency for model training, which is the core intuition of DSL. Therefore, it is not necessary to test DSL when \textit{T} = 1. As for CDSL, even if there is only one target in the training sequence, the conditional set likelihood can be treated as the probability of selecting a DPP-distributed item set (previous items and a target) given the previous items. This means that only \textit{sequence dependency} is considered when \textit{T} = 1, and both \textit{sequence dependency} and \textit{targets dependency} are taken into account when \textit{T} > 1. 
The main observations from Table 2-4 are as follows:
\begin{itemize}[leftmargin=*]
\vspace{-2mm}
\item Aligning these three tables, we can see that deploying DSL or CDSL on the state-of-the-art sequential recommendation models significantly improves recommendation quality (recall and NDCG) \textit{w.r.t.} different \textit{Top-N} on different datasets. This suggests that set likelihood-based objective functions that consider sequence dependency and targets dependency are more adaptable to sequential recommendation tasks than commonly used ranking-based loss functions.
\item In only one case --- MARank model on Beauty dataset when \textit{T} = 3, the value of \textit{trade-off} metric F@10 is slightly higher than DSL and CDSL. Under two settings (ML-1M dataset when \textit{T} = 3 and Anime dataset when \textit{T} = 3), SASRec outperforms the reworked DSL or CDSL version \textit{w.r.t.} trade-off metric. In addition to these three cases, DSL and CDSL achieve better trade-off performance than the original loss functions of baselines. In some cases, the reworked models outperform baselines in F-score by a large margin (over 20\%), \textit{e.g.}, reworked MARank on Anime when $T$ = 3. These demonstrate that DSL and CDSL can better handle the balance between quality and diversity by DPP kernel decomposition, \textit{i.e.}, facilitating recommending relevant and diverse results.
\item When the original models achieve better diversification (CC metric) than their reworked versions, \textit{e.g.}, experiments on Anime dataset ($T$ = 1) in Table 2, their corresponding quality performances (Recall and NDCG at the same top $N$) are clearly inadequate. This observation shows that state-of-the-art baselines sacrifice recommendation relevance for diversity (\textit{i.e.}, sub-optimally trade-off between quality and diversity).
\item In those few instances that baseline obtains better recommendation quality than DSL or CDSL in three tables, they are all at top-5 or top-10 (\textit{e.g.}, Re@10 on ML-1M when $T$ = 1 in Table 2). This indicates that in a few cases (only 5 values in Table 2-4), DSL and CDSL may focus more on providing the 'really' relevant (\textit{e.g.}, top-3) items with a high probability of being drawn as a DPP set but ignore the relatively less relevant ones in a list.
\item By comparing the performances under different settings of target length \textit{T} in Table 1, \textit{T} = 3 setting generally yields the most significant improvement compared to the baseline Caser. The reason might be that as the number of target items increases (\textit{e.g.}, \textit{T} = 5), noisy or confusing dependencies between the targets is considered, as thus affecting recommendations. In view of this phenomenon, Table 3 and 4 only present the results under \textit{T} = 1 and \textit{T} = 3 settings. 
\item Among three experimental datasets in Table 2-4, DSL and CDSL, in general, achieve the most significant improvement on Beauty dataset compared to the corresponding baseline. We think this is because on shopping Web platforms, users tend to search or buy products (items) that have tight correlations over a period. This verifies the importance of dependencies in the sequence or among targets to some extent, because dependencies are considered in the proposed loss functions. 
\item Compared to DSL in three tables, CDSL in general achieves better overall performance (F@N) when $T$ > 1. This further verifies the importance of considering sequence dependency and targets dependency in the loss function, as CDSL considers both of them. Although CDSL cannot beat DSL in most cases with respect to quality metrics, it has a distinct advantage in diversification (CC@N). This finding tells us that taking previous actions into account (\textit{i.e.}, increasing the size of the ground set $\mathcal{Y}^{(u,t)}$) is likely to make the decomposition of user-sequence specified DPP kernel care more about diversity than quality, thus leading CDSL to achieve better diversity performance. This can also be treated as an advantage of CDSL, because it is the distinct diversity result that contributes to the better overall performance. 
\end{itemize} 

Among all observations, there is no comparison between the performances of any two methods under the same setting but in different tables. This is because our purpose of presenting Table 2-4 is to demonstrate the superiority of DSL and CDSL through analyzing the performance improvement brought by using DSL or CDSL compared to the corresponding baseline with its original loss function under the same setting. Therefore, there is no need to compare the results of different tables. In the design of Caser, $T$ targets are considered in loss calculation, so we can directly feed training instances of $L$ previous items and $T$ targets into its \textit{binary cross-entropy loss}. Although the loss functions in MARank and SASRec are designed only for next one target, we still can deploy DSL or CDSL in them by modifying the training sequences and expand the original losses that consider one target to comparing $T$ targets (Equation 1 and Equation 3) when $T>1$. As the architecture of baselines is not altered, performance comparisons are still fair. Essentially, we only compare the loss functions.  

We use Figure 2 to investigate the effect of negative sample length $Z$. It shows that for all models (Caser and its reworked versions --- DSL and CDSL) on ML-1M and Beauty under the setting of $T = 2$ and $L=5$, with the increase of $Z$ the overall trend is that the NDCG@5 performance grows rapidly at first and then gradually drops. For all experiments reported in Table 2-4, $Z$ is set according to the target length $T$ like most recommendation models do, instead of searching the optimal value of $Z$ by tuning $Z$. This helps to keep the comparisons fair and the training efficient. In most cases, DSL and CDSL perform better than Caser and the improvement over Caser increases as the value of $Z$ gets larger. When $Z$ equals 1, the performance of CDSL on ML-1M is not better than Caser. This may be because there is not enough pattern/correlation for learning a competent DPP probability measure, when conditioning on 5 previous items for drawing a DPP set with 7 items ($L+T$) that almost equals the item number of the ground set ($L+T+Z=8$). These observations indicate that negative samples are not only important for the original loss function but also for DSL and CDSL. 

\begin{figure*}
\centering
    \subfigure[lr=0.0005]{
    \includegraphics[width=0.32\linewidth]{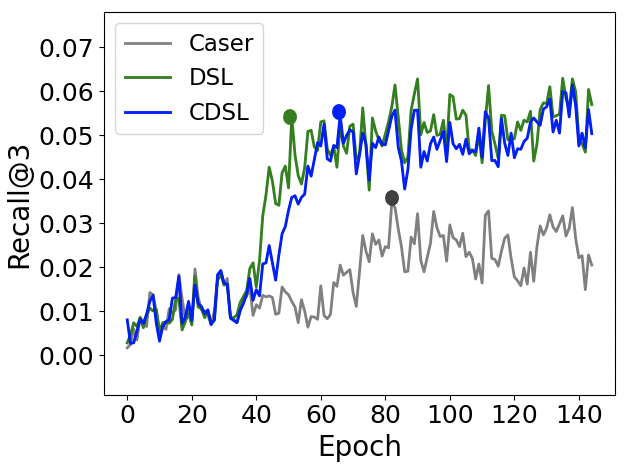}}
    \subfigure[lr=0.001]{
    \includegraphics[width=0.32\linewidth]{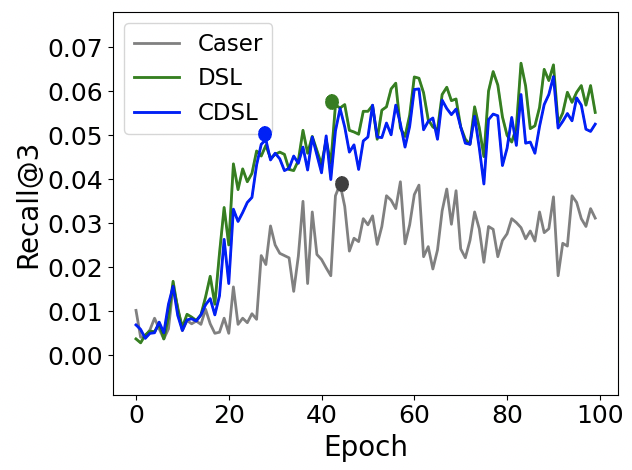}}
    \subfigure[lr=0.0015]{
    \includegraphics[width=0.32\linewidth]{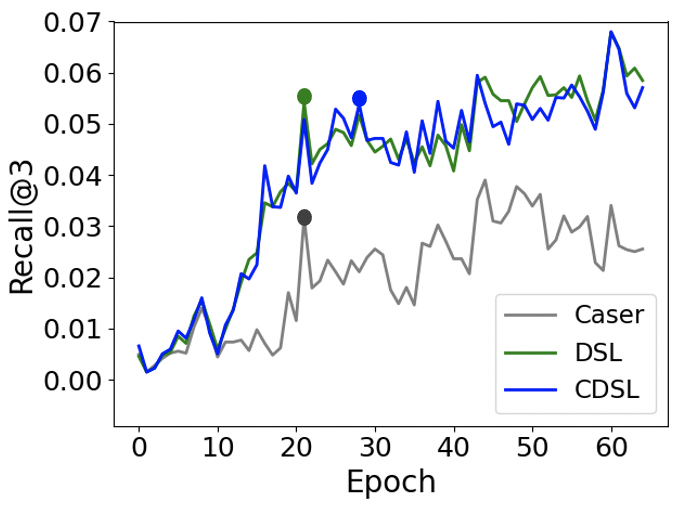}}
\vspace{-1mm}
\caption{Test performance of each epoch with different learning rate on Beauty.}
\vspace{-3mm}
\end{figure*}

\begin{figure*}
\centering
    \subfigure[lr=0.0005]{
    \includegraphics[width=0.32\linewidth]{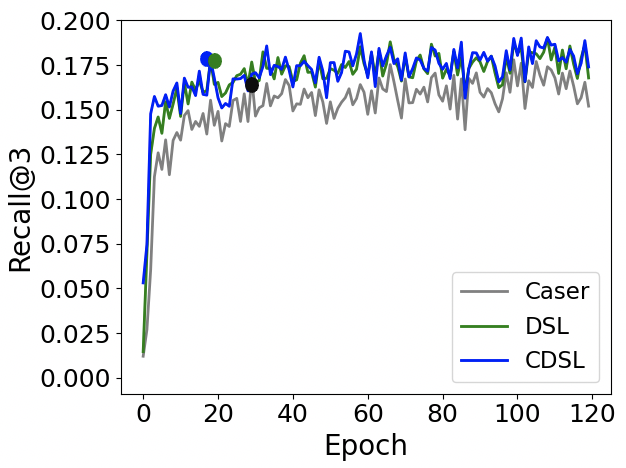}}
    \subfigure[lr=0.001]{
    \includegraphics[width=0.32\linewidth]{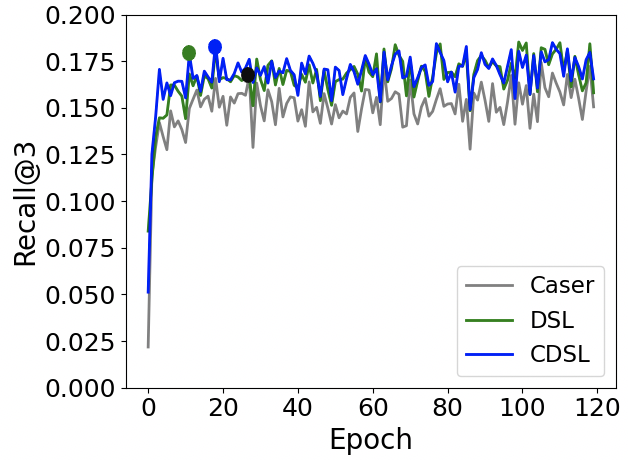}}
    \subfigure[lr=0.0015]{
    \includegraphics[width=0.32\linewidth]{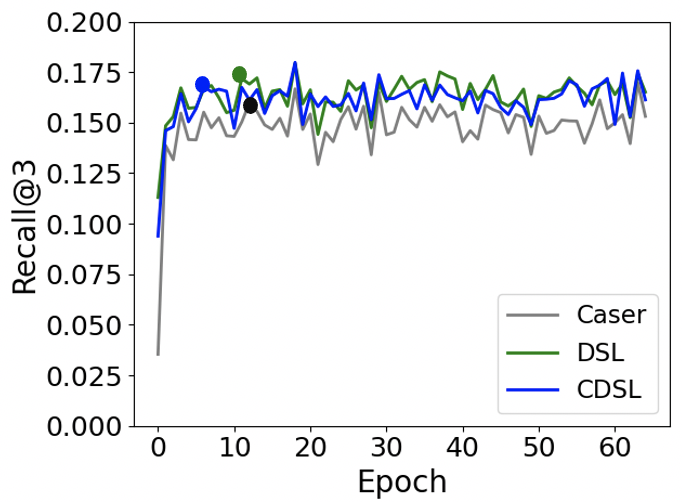}}
\vspace{-1mm}
\caption{Test performance of each epoch with different learning rate on Anime.}
\vspace{-3mm}
\end{figure*}

Figure 3 shows the test performance (Recall@3) of each training epoch of three models with different learning rates (\texttt{lr}) on Beauty. Figure 4 presents the same type of performance comparison on Anime. The value of 0.001 is the default setting of \texttt{lr} for the Caser baseline, which is also used for DSL and CDSL versions, with the same setting ($T = 3$, $L = 5$, and $Z = 3$) for all experiments. The solid circle on a performance curve of each subfigure indicates the best performance before the early stopping. In Figure 3 and 4, we use Recall@3 metric to judge whether the early stopping strategy is triggered instead of using NDCG@5 as mentioned above. The solid circle has the same color as the corresponding model's performance curve. 
Two main observations can be made: \textit{i)} the proposed loss functions achieve better performance almost at any epoch in six subfigures (three \texttt{lr} settings on two datasets) compared to the baseline, indicating the superiority of our loss functions in a more detailed way; \textit{ii)} in general, the proposed loss functions prefer a smaller learning rate compared to the original loss, and they learn faster (\textit{i.e.}, trigger the early stopping earlier) in most cases. As the reworked models of other baselines share similar trends with Caser on different datasets, other subfigures with different settings are not presented. 

\begin{table*}[tp]
\centering
  \fontsize{8.3}{10.}\selectfont
  \caption{Training efficiency comparison on two datasets.}
  \label{tab:performance_comparison_efficiency}
    \begin{tabular}{|c|c|ccc|cccc|cccc|}
    \hline
    \multicolumn{2}{|c|}{\multirow{2}{*}{Datasets}}&\multicolumn{3}{c|}{MARank}&\multicolumn{4}{c|}{DSL}&\multicolumn{4}{c|}{CDSL}\\\cline{3-13}
    \multicolumn{2}{|c|}{} & 
    \texttt{Time}&\#\texttt{Epochs}&\texttt{Total} &\texttt{Time}&\#\texttt{Epochs}&\texttt{Total}&\texttt{pct.}& \texttt{Time}&\#\texttt{Epochs}&\texttt{Total}&\texttt{pct.}
    \cr\hline
    \multirow{2}{*}{Anime} 
    & \textit{T}=1&22.03&35&771.05&-&-&-&-&48.67&17&827.39&7.30\% \\ \cline{2-13}
    & \textit{T}=3&25.37&38&964.06&52.72&19&1001.68&3.90\%&55.15&19&1047.85&8.69\% \\
    \hline
    \hline
    \multirow{2}{*}{Beauty} 
    & \textit{T}=1&1.02&40&40.80&-&-&-&-&1.88&23&43.24& 5.98\% \\\cline{2-13}
    & \textit{T}=3&1.06&42&44.52&1.94&23&44.62& 0.22\%&1.98&25&49.50& 11.19\% \\
    \hline
    \end{tabular}
\end{table*}

Table 5 shows the training efficiency comparison between MARank (using pairwise loss) and reworked models with our loss functions on two datasets under different settings of target length. In the table, \texttt{Time} means average training time per epoch (in seconds), and \#\texttt{Epochs} represents the number of epochs needed to converge (achieve best NDCG@5 performance before early stopping). \texttt{Total} records the total time used in a training process \texttt{Time} $\times$ \#\texttt{Epochs}. Besides, \texttt{pct.} denotes the \textit{percent change} between our reworked models and the baseline. All experiments in Table 5 are conducted using a single Tesla K40 GPU under the same setting. We can see that the reworked models with DSL or CDSL take more time per epoch than MARank, but converge faster (fewer epochs are needed in training process) and perform better (shown in Table 3). This observation is also made in Figure 3 and 4. In general, DSL and CDSL only need around 10\% more time to coverage than that of MARank, but significant improvements are achieved (as shown in Table 2-4). In our experiments, when we compare the training efficiency based on Caser or SASRec (using pointwise ranking), similar observations to those of Table 5 are made. Therefore, comparisons based on other baselines and datasets are not presented. 

From the analysis of Table 2-4 and Figure 2, we can draw a clear conclusion about how to make a choice between DSL and CDSL in certain circumstances. If a sequential recommender system emphasizes relevance and training efficiency, DSL is a better choice. On the other hand, in the case where overall performance between quality and diversity is the main concern (\textit{e.g.}, diversity-promoting recommendation \cite{wu2019pd, liang2021recommending}) and training efficiency is not the primary factor, choosing CDSL is more advisable. 
\vspace{-3mm}
\section{RELATED WORK}
Three lines of work are closely related to ours, which are listed as follows.
\vspace{2mm}
\\ \textbf{Sequential Recommendation.}
The main factor that sets the sequential recommendation apart from conventional recommender systems (\textit{e.g.}, \textit{Top-N} recommendation and collaborative filtering) is the consideration of sequential dependency. To capture sequential dependency for the user interaction sequence completion, Markov Chains based \cite{rendle2010factorizing, he2016fusing} and Recurrent Neural Networks based \cite{tang2018personalized, kang2018self, sun2019bert4rec} methods have dominated the literature and real-word applications. In this paper, we propose DSL and CDSL loss functions targeting next target(s) estimation. The next-item recommendation is a common task \cite{yuan2019simple, tang2018personalized} that calculate a loss based on next one target in the training process. The 'next items' in this paper actually means the next $T$ targets for estimation in training. By preparing user sequences with $L$ previous items and $T$ targets for training, we aim to introduce the targets dependency into the loss formulation, and then facilitate recommendation models providing users with a set of relevant results. In this setting, next items recommendation is not the same as next-basket recommendation that predicts the next basket for a user based on a set of sequential baskets (\textit{e.g.}, shopping baskets or transactions), as the basket length (number of products/items in basket) is not fixed. The session-based recommendation \cite{hidasi2015session, li2017neural} is a kind of non-personalized sequential recommendation, which uses sequential session data without user profile/identifier for recommendation. We believe that our loss functions are also adaptive to next-basket and session-based recommendations (both consider the temporal dependency), which will be explored in the future. 
\vspace{2mm}
\\ \textbf{Loss Functions in Sequential Recommendation.}
In the sequential recommendation literature, two types of loss functions are widely employed, \textit{i.e.}, binary cross-entropy and pairwise ranking (BPR). Many different variations are also developed based on these two loss functions, \textit{e.g.}, softmax cross-entropy \cite{covington2016deep} and pairwise hinge loss \cite{hsieh2017collaborative}. 
BPR is usually limited to having only one target and one negative sample at each time step \cite{tang2018personalized, yu2019multi, socher2013reasoning}, but some studies have expanded the BPR formulation to a set-wise Bayesian approach \cite{wang2020setrank, rendle2010factorizing} to consider more comparisons. 
The seminal session-based recommendation model GRU4Rec \cite{hidasi2015session} proposes the TOP1 loss which is similar to the BPR loss and an improved version GRU4Rec$^+$ \cite{hidasi2018recurrent} further boosts the recommendation performance using a special case of binary cross-entropy, \textit{i.e.}, categorical cross-entropy. 
Recently, inspired by the widely used contrastive loss \cite{hadsell2006dimensionality, yang2019adaptive} in the computer vision tasks, a line of work \cite{xie2020contrastive, xie2021adversarial} employs the contrastive loss in sequential recommendation models, which is defined similarly to the softmax cross entropy loss. The nature of the above-mentioned loss functions is still the relevance-based ranking \cite{hidasi2018recurrent}, and the dependencies in sequence and among targets are ignored. 
In addition, the temporal dependency is complex \cite{wu2019dual}, merely considering the dependency of items' ranking positions in a list (\textit{e.g.}, list-wise ranking) is not enough \cite{chen2021set2setrank}. Therefore, we treat the sequential recommendation as a task of set selection using the elegant probabilistic model --- DPP, instead of directly comparing the rankings of items. 
\vspace{2mm}
\\ \textbf{Determinantal Point Process.}
Employing DPP to diversify recommendations is a sensible choice, as DPP can measure the diversity of a set by describing the probability for all subsets of entire items. However, the main application of DPP that uses the maximum a posteriori (MAP) to generate diverse elements is NP-hard and thus computationally expensive \cite{liang2021recommending}. With the development of a novel algorithm that accelerates MAP inference for DPP based on the fast inference method \cite{chen2018fast}, some practical DPP-based diversity-promoting methods \cite{wu2019pd, gan2020enhancing} are proposed to facilitate traditional recommendations providing diverse and relevant results. In addition to using MAP inference for item generation, there are some works \cite{warlop2019tensorized, wilhelm2018practical} proposed to parameterize the DPP kernel on entire items and then learn the kernel matrix as item representations for recommendation.
This work is quite different from previous studies, as we directly propose two types of loss functions (DSL and CDSL) based on the likelihoods of DPP set. By considering special dependency correlations in temporal sequences, DSL and CDSL can fit naturally into different sequential recommendation pipelines.

\section{CONCLUSION}
In this work, we proposed two novel DPP set likelihood-based loss functions (DSL and CDSL) aiming at sequential recommendations, which enlighten a new perspective (set selection) to formulate the temporal training objectives. As such, the \textit{sequence dependency} and \textit{targets dependency} are considered in the loss formulations. In addition, through pre-learning a diverse kernel from observed browsing history and using the quality \textit{vs.} diversity decomposition of user-sequence specified DPP kernel, the proposed loss functions are pushed to be diversity-aware. By simply replacing the original loss of state-of-the-art sequential recommendation models with our DSL or CDSL objective functions, significant improvements can be received, which demonstrates the superiority and flexibility of the proposed loss functions. Experimental results help us analyze the different advantages of the two loss functions and provide insight into making selection between them in different circumstances. 
DSL and CDSL can also be applied to other temporal sequence related tasks, such as clinical events prediction \cite{hu2019sets2sets, choi2016doctor} and next-basket recommendation, which will be explored in the future. 
\vspace{3mm} \\
\textbf{Acknowledgement:} This research is supported in part by the Australian Research Council Project DP180101985.
%%
%% The next two lines define the bibliography style to be used, and
%% the bibliography file.
\normalem
\bibliographystyle{ACM-Reference-Format}
\bibliography{reference}

%%% -*-BibTeX-*-
%%% Do NOT edit. File created by BibTeX with style
%%% ACM-Reference-Format-Journals [18-Jan-2012].

\begin{thebibliography}{52}

%%% ====================================================================
%%% NOTE TO THE USER: you can override these defaults by providing
%%% customized versions of any of these macros before the \bibliography
%%% command.  Each of them MUST provide its own final punctuation,
%%% except for \shownote{}, \showDOI{}, and \showURL{}.  The latter two
%%% do not use final punctuation, in order to avoid confusing it with
%%% the Web address.
%%%
%%% To suppress output of a particular field, define its macro to expand
%%% to an empty string, or better, \unskip, like this:
%%%
%%% \newcommand{\showDOI}[1]{\unskip}   % LaTeX syntax
%%%
%%% \def \showDOI #1{\unskip}           % plain TeX syntax
%%%
%%% ====================================================================

\ifx \showCODEN    \undefined \def \showCODEN     #1{\unskip}     \fi
\ifx \showDOI      \undefined \def \showDOI       #1{#1}\fi
\ifx \showISBNx    \undefined \def \showISBNx     #1{\unskip}     \fi
\ifx \showISBNxiii \undefined \def \showISBNxiii  #1{\unskip}     \fi
\ifx \showISSN     \undefined \def \showISSN      #1{\unskip}     \fi
\ifx \showLCCN     \undefined \def \showLCCN      #1{\unskip}     \fi
\ifx \shownote     \undefined \def \shownote      #1{#1}          \fi
\ifx \showarticletitle \undefined \def \showarticletitle #1{#1}   \fi
\ifx \showURL      \undefined \def \showURL       {\relax}        \fi
% The following commands are used for tagged output and should be
% invisible to TeX
\providecommand\bibfield[2]{#2}
\providecommand\bibinfo[2]{#2}
\providecommand\natexlab[1]{#1}
\providecommand\showeprint[2][]{arXiv:#2}

\bibitem[\protect\citeauthoryear{Borodin and Rains}{Borodin and Rains}{2005}]%
        {borodin2005eynard}
\bibfield{author}{\bibinfo{person}{Alexei Borodin} {and}
  \bibinfo{person}{Eric~M Rains}.} \bibinfo{year}{2005}\natexlab{}.
\newblock \showarticletitle{Eynard--Mehta theorem, Schur process, and their
  Pfaffian analogs}.
\newblock \bibinfo{journal}{\emph{Journal of statistical physics}}
  \bibinfo{volume}{121}, \bibinfo{number}{3} (\bibinfo{year}{2005}),
  \bibinfo{pages}{291--317}.
\newblock


\bibitem[\protect\citeauthoryear{Cao, Qin, Liu, Tsai, and Li}{Cao
  et~al\mbox{.}}{2007}]%
        {cao2007learning}
\bibfield{author}{\bibinfo{person}{Zhe Cao}, \bibinfo{person}{Tao Qin},
  \bibinfo{person}{Tie-Yan Liu}, \bibinfo{person}{Ming-Feng Tsai}, {and}
  \bibinfo{person}{Hang Li}.} \bibinfo{year}{2007}\natexlab{}.
\newblock \showarticletitle{Learning to rank: from pairwise approach to
  listwise approach}. In \bibinfo{booktitle}{\emph{Proceedings of the 24th
  international conference on Machine learning}}. \bibinfo{pages}{129--136}.
\newblock


\bibitem[\protect\citeauthoryear{Chen, Wu, Zhang, Hong, and Wang}{Chen
  et~al\mbox{.}}{2021}]%
        {chen2021set2setrank}
\bibfield{author}{\bibinfo{person}{Lei Chen}, \bibinfo{person}{Le Wu},
  \bibinfo{person}{Kun Zhang}, \bibinfo{person}{Richang Hong}, {and}
  \bibinfo{person}{Meng Wang}.} \bibinfo{year}{2021}\natexlab{}.
\newblock \showarticletitle{Set2setRank: Collaborative Set to Set Ranking for
  Implicit Feedback based Recommendation}.
\newblock \bibinfo{journal}{\emph{arXiv preprint arXiv:2105.07377}}
  (\bibinfo{year}{2021}).
\newblock


\bibitem[\protect\citeauthoryear{Chen, Zhang, and Zhou}{Chen
  et~al\mbox{.}}{2018}]%
        {chen2018fast}
\bibfield{author}{\bibinfo{person}{Laming Chen}, \bibinfo{person}{Guoxin
  Zhang}, {and} \bibinfo{person}{Hanning Zhou}.}
  \bibinfo{year}{2018}\natexlab{}.
\newblock \showarticletitle{Fast greedy map inference for determinantal point
  process to improve recommendation diversity}. In
  \bibinfo{booktitle}{\emph{Proceedings of the 32nd International Conference on
  Neural Information Processing Systems}}. \bibinfo{pages}{5627--5638}.
\newblock


\bibitem[\protect\citeauthoryear{Cheng, Wang, Ma, Sun, and Xiong}{Cheng
  et~al\mbox{.}}{2017}]%
        {cheng2017learning}
\bibfield{author}{\bibinfo{person}{Peizhe Cheng}, \bibinfo{person}{Shuaiqiang
  Wang}, \bibinfo{person}{Jun Ma}, \bibinfo{person}{Jiankai Sun}, {and}
  \bibinfo{person}{Hui Xiong}.} \bibinfo{year}{2017}\natexlab{}.
\newblock \showarticletitle{Learning to recommend accurate and diverse items}.
  In \bibinfo{booktitle}{\emph{Proceedings of the 26th international conference
  on World Wide Web}}. \bibinfo{pages}{183--192}.
\newblock


\bibitem[\protect\citeauthoryear{Choi, Bahadori, Schuetz, Stewart, and
  Sun}{Choi et~al\mbox{.}}{2016}]%
        {choi2016doctor}
\bibfield{author}{\bibinfo{person}{Edward Choi}, \bibinfo{person}{Mohammad~Taha
  Bahadori}, \bibinfo{person}{Andy Schuetz}, \bibinfo{person}{Walter~F
  Stewart}, {and} \bibinfo{person}{Jimeng Sun}.}
  \bibinfo{year}{2016}\natexlab{}.
\newblock \showarticletitle{Doctor ai: Predicting clinical events via recurrent
  neural networks}. In \bibinfo{booktitle}{\emph{Machine learning for
  healthcare conference}}. PMLR, \bibinfo{pages}{301--318}.
\newblock


\bibitem[\protect\citeauthoryear{Covington, Adams, and Sargin}{Covington
  et~al\mbox{.}}{2016}]%
        {covington2016deep}
\bibfield{author}{\bibinfo{person}{Paul Covington}, \bibinfo{person}{Jay
  Adams}, {and} \bibinfo{person}{Emre Sargin}.}
  \bibinfo{year}{2016}\natexlab{}.
\newblock \showarticletitle{Deep neural networks for youtube recommendations}.
  In \bibinfo{booktitle}{\emph{Proceedings of the 10th ACM conference on
  recommender systems}}. \bibinfo{pages}{191--198}.
\newblock


\bibitem[\protect\citeauthoryear{Deshpande and Karypis}{Deshpande and
  Karypis}{2004}]%
        {deshpande2004item}
\bibfield{author}{\bibinfo{person}{Mukund Deshpande} {and}
  \bibinfo{person}{George Karypis}.} \bibinfo{year}{2004}\natexlab{}.
\newblock \showarticletitle{Item-based top-n recommendation algorithms}.
\newblock \bibinfo{journal}{\emph{ACM Transactions on Information Systems
  (TOIS)}} \bibinfo{volume}{22}, \bibinfo{number}{1} (\bibinfo{year}{2004}),
  \bibinfo{pages}{143--177}.
\newblock


\bibitem[\protect\citeauthoryear{Fang, Zhang, Shu, and Guo}{Fang
  et~al\mbox{.}}{2020}]%
        {fang2020deep}
\bibfield{author}{\bibinfo{person}{Hui Fang}, \bibinfo{person}{Danning Zhang},
  \bibinfo{person}{Yiheng Shu}, {and} \bibinfo{person}{Guibing Guo}.}
  \bibinfo{year}{2020}\natexlab{}.
\newblock \showarticletitle{Deep learning for sequential recommendation:
  Algorithms, influential factors, and evaluations}.
\newblock \bibinfo{journal}{\emph{ACM Transactions on Information Systems
  (TOIS)}} \bibinfo{volume}{39}, \bibinfo{number}{1} (\bibinfo{year}{2020}),
  \bibinfo{pages}{1--42}.
\newblock


\bibitem[\protect\citeauthoryear{Gan, Nurbakova, Laporte, and Calabretto}{Gan
  et~al\mbox{.}}{2020}]%
        {gan2020enhancing}
\bibfield{author}{\bibinfo{person}{Lu Gan}, \bibinfo{person}{Diana Nurbakova},
  \bibinfo{person}{L{\'e}a Laporte}, {and} \bibinfo{person}{Sylvie
  Calabretto}.} \bibinfo{year}{2020}\natexlab{}.
\newblock \showarticletitle{Enhancing recommendation diversity using
  determinantal point processes on knowledge graphs}. In
  \bibinfo{booktitle}{\emph{Proceedings of the 43rd International ACM SIGIR
  Conference on Research and Development in Information Retrieval}}.
  \bibinfo{pages}{2001--2004}.
\newblock


\bibitem[\protect\citeauthoryear{Gartrell, Brunel, Dohmatob, and
  Krichene}{Gartrell et~al\mbox{.}}{2019}]%
        {gartrell2019learning}
\bibfield{author}{\bibinfo{person}{Mike Gartrell},
  \bibinfo{person}{Victor-Emmanuel Brunel}, \bibinfo{person}{Elvis Dohmatob},
  {and} \bibinfo{person}{Syrine Krichene}.} \bibinfo{year}{2019}\natexlab{}.
\newblock \showarticletitle{Learning nonsymmetric determinantal point
  processes}.
\newblock \bibinfo{journal}{\emph{arXiv preprint arXiv:1905.12962}}
  (\bibinfo{year}{2019}).
\newblock


\bibitem[\protect\citeauthoryear{Gartrell, Paquet, and Koenigstein}{Gartrell
  et~al\mbox{.}}{2017}]%
        {gartrell2017low}
\bibfield{author}{\bibinfo{person}{Mike Gartrell}, \bibinfo{person}{Ulrich
  Paquet}, {and} \bibinfo{person}{Noam Koenigstein}.}
  \bibinfo{year}{2017}\natexlab{}.
\newblock \showarticletitle{Low-rank factorization of determinantal point
  processes}. In \bibinfo{booktitle}{\emph{Thirty-First AAAI Conference on
  Artificial Intelligence}}.
\newblock


\bibitem[\protect\citeauthoryear{Hadsell, Chopra, and LeCun}{Hadsell
  et~al\mbox{.}}{2006}]%
        {hadsell2006dimensionality}
\bibfield{author}{\bibinfo{person}{Raia Hadsell}, \bibinfo{person}{Sumit
  Chopra}, {and} \bibinfo{person}{Yann LeCun}.}
  \bibinfo{year}{2006}\natexlab{}.
\newblock \showarticletitle{Dimensionality reduction by learning an invariant
  mapping}. In \bibinfo{booktitle}{\emph{2006 IEEE Computer Society Conference
  on Computer Vision and Pattern Recognition (CVPR'06)}},
  Vol.~\bibinfo{volume}{2}. IEEE, \bibinfo{pages}{1735--1742}.
\newblock


\bibitem[\protect\citeauthoryear{He, Kang, and McAuley}{He
  et~al\mbox{.}}{2017a}]%
        {he2017translation}
\bibfield{author}{\bibinfo{person}{Ruining He}, \bibinfo{person}{Wang-Cheng
  Kang}, {and} \bibinfo{person}{Julian McAuley}.}
  \bibinfo{year}{2017}\natexlab{a}.
\newblock \showarticletitle{Translation-based recommendation}. In
  \bibinfo{booktitle}{\emph{Proceedings of the eleventh ACM conference on
  recommender systems}}. \bibinfo{pages}{161--169}.
\newblock


\bibitem[\protect\citeauthoryear{He and McAuley}{He and McAuley}{2016}]%
        {he2016fusing}
\bibfield{author}{\bibinfo{person}{Ruining He} {and} \bibinfo{person}{Julian
  McAuley}.} \bibinfo{year}{2016}\natexlab{}.
\newblock \showarticletitle{Fusing similarity models with markov chains for
  sparse sequential recommendation}. In \bibinfo{booktitle}{\emph{2016 IEEE
  16th International Conference on Data Mining (ICDM)}}. IEEE,
  \bibinfo{pages}{191--200}.
\newblock


\bibitem[\protect\citeauthoryear{He, Liao, Zhang, Nie, Hu, and Chua}{He
  et~al\mbox{.}}{2017b}]%
        {he2017neural}
\bibfield{author}{\bibinfo{person}{Xiangnan He}, \bibinfo{person}{Lizi Liao},
  \bibinfo{person}{Hanwang Zhang}, \bibinfo{person}{Liqiang Nie},
  \bibinfo{person}{Xia Hu}, {and} \bibinfo{person}{Tat-Seng Chua}.}
  \bibinfo{year}{2017}\natexlab{b}.
\newblock \showarticletitle{Neural collaborative filtering}. In
  \bibinfo{booktitle}{\emph{Proceedings of the 26th international conference on
  world wide web}}. \bibinfo{pages}{173--182}.
\newblock


\bibitem[\protect\citeauthoryear{Hidasi and Karatzoglou}{Hidasi and
  Karatzoglou}{2018}]%
        {hidasi2018recurrent}
\bibfield{author}{\bibinfo{person}{Bal{\'a}zs Hidasi} {and}
  \bibinfo{person}{Alexandros Karatzoglou}.} \bibinfo{year}{2018}\natexlab{}.
\newblock \showarticletitle{Recurrent neural networks with top-k gains for
  session-based recommendations}. In \bibinfo{booktitle}{\emph{Proceedings of
  the 27th ACM international conference on information and knowledge
  management}}. \bibinfo{pages}{843--852}.
\newblock


\bibitem[\protect\citeauthoryear{Hidasi, Karatzoglou, Baltrunas, and
  Tikk}{Hidasi et~al\mbox{.}}{2015}]%
        {hidasi2015session}
\bibfield{author}{\bibinfo{person}{Bal{\'a}zs Hidasi},
  \bibinfo{person}{Alexandros Karatzoglou}, \bibinfo{person}{Linas Baltrunas},
  {and} \bibinfo{person}{Domonkos Tikk}.} \bibinfo{year}{2015}\natexlab{}.
\newblock \showarticletitle{Session-based recommendations with recurrent neural
  networks}.
\newblock \bibinfo{journal}{\emph{arXiv preprint arXiv:1511.06939}}
  (\bibinfo{year}{2015}).
\newblock


\bibitem[\protect\citeauthoryear{Hsieh, Yang, Cui, Lin, Belongie, and
  Estrin}{Hsieh et~al\mbox{.}}{2017}]%
        {hsieh2017collaborative}
\bibfield{author}{\bibinfo{person}{Cheng-Kang Hsieh}, \bibinfo{person}{Longqi
  Yang}, \bibinfo{person}{Yin Cui}, \bibinfo{person}{Tsung-Yi Lin},
  \bibinfo{person}{Serge Belongie}, {and} \bibinfo{person}{Deborah Estrin}.}
  \bibinfo{year}{2017}\natexlab{}.
\newblock \showarticletitle{Collaborative metric learning}. In
  \bibinfo{booktitle}{\emph{Proceedings of the 26th international conference on
  world wide web}}. \bibinfo{pages}{193--201}.
\newblock


\bibitem[\protect\citeauthoryear{Hu and He}{Hu and He}{2019}]%
        {hu2019sets2sets}
\bibfield{author}{\bibinfo{person}{Haoji Hu} {and} \bibinfo{person}{Xiangnan
  He}.} \bibinfo{year}{2019}\natexlab{}.
\newblock \showarticletitle{Sets2sets: Learning from sequential sets with
  neural networks}. In \bibinfo{booktitle}{\emph{Proceedings of the 25th ACM
  SIGKDD International Conference on Knowledge Discovery \& Data Mining}}.
  \bibinfo{pages}{1491--1499}.
\newblock


\bibitem[\protect\citeauthoryear{Hu, Koren, and Volinsky}{Hu
  et~al\mbox{.}}{2008}]%
        {hu2008collaborative}
\bibfield{author}{\bibinfo{person}{Yifan Hu}, \bibinfo{person}{Yehuda Koren},
  {and} \bibinfo{person}{Chris Volinsky}.} \bibinfo{year}{2008}\natexlab{}.
\newblock \showarticletitle{Collaborative filtering for implicit feedback
  datasets}. In \bibinfo{booktitle}{\emph{2008 Eighth IEEE International
  Conference on Data Mining}}. Ieee, \bibinfo{pages}{263--272}.
\newblock


\bibitem[\protect\citeauthoryear{Kang and McAuley}{Kang and McAuley}{2018}]%
        {kang2018self}
\bibfield{author}{\bibinfo{person}{Wang-Cheng Kang} {and}
  \bibinfo{person}{Julian McAuley}.} \bibinfo{year}{2018}\natexlab{}.
\newblock \showarticletitle{Self-attentive sequential recommendation}. In
  \bibinfo{booktitle}{\emph{2018 IEEE International Conference on Data Mining
  (ICDM)}}. IEEE, \bibinfo{pages}{197--206}.
\newblock


\bibitem[\protect\citeauthoryear{Kulesza and Taskar}{Kulesza and
  Taskar}{2011}]%
        {kulesza2011k}
\bibfield{author}{\bibinfo{person}{Alex Kulesza} {and} \bibinfo{person}{Ben
  Taskar}.} \bibinfo{year}{2011}\natexlab{}.
\newblock \showarticletitle{k-DPPs: Fixed-size determinantal point processes}.
  In \bibinfo{booktitle}{\emph{ICML}}.
\newblock


\bibitem[\protect\citeauthoryear{Kulesza and Taskar}{Kulesza and
  Taskar}{2012}]%
        {kulesza2012determinantal}
\bibfield{author}{\bibinfo{person}{Alex Kulesza} {and} \bibinfo{person}{Ben
  Taskar}.} \bibinfo{year}{2012}\natexlab{}.
\newblock \showarticletitle{Determinantal point processes for machine
  learning}.
\newblock \bibinfo{journal}{\emph{arXiv preprint arXiv:1207.6083}}
  (\bibinfo{year}{2012}).
\newblock


\bibitem[\protect\citeauthoryear{Li, Ren, Chen, Ren, Lian, and Ma}{Li
  et~al\mbox{.}}{2017}]%
        {li2017neural}
\bibfield{author}{\bibinfo{person}{Jing Li}, \bibinfo{person}{Pengjie Ren},
  \bibinfo{person}{Zhumin Chen}, \bibinfo{person}{Zhaochun Ren},
  \bibinfo{person}{Tao Lian}, {and} \bibinfo{person}{Jun Ma}.}
  \bibinfo{year}{2017}\natexlab{}.
\newblock \showarticletitle{Neural attentive session-based recommendation}. In
  \bibinfo{booktitle}{\emph{Proceedings of the 2017 ACM on Conference on
  Information and Knowledge Management}}. \bibinfo{pages}{1419--1428}.
\newblock


\bibitem[\protect\citeauthoryear{Li, Wang, and McAuley}{Li
  et~al\mbox{.}}{2020}]%
        {li2020time}
\bibfield{author}{\bibinfo{person}{Jiacheng Li}, \bibinfo{person}{Yujie Wang},
  {and} \bibinfo{person}{Julian McAuley}.} \bibinfo{year}{2020}\natexlab{}.
\newblock \showarticletitle{Time interval aware self-attention for sequential
  recommendation}. In \bibinfo{booktitle}{\emph{Proceedings of the 13th
  international conference on web search and data mining}}.
  \bibinfo{pages}{322--330}.
\newblock


\bibitem[\protect\citeauthoryear{Li, Chen, Zhang, and Yin}{Li
  et~al\mbox{.}}{2021}]%
        {li2021lightweight}
\bibfield{author}{\bibinfo{person}{Yang Li}, \bibinfo{person}{Tong Chen},
  \bibinfo{person}{Peng-Fei Zhang}, {and} \bibinfo{person}{Hongzhi Yin}.}
  \bibinfo{year}{2021}\natexlab{}.
\newblock \showarticletitle{Lightweight Self-Attentive Sequential
  Recommendation}. In \bibinfo{booktitle}{\emph{Proceedings of the 30th ACM
  International Conference on Information \& Knowledge Management}}.
  \bibinfo{pages}{967--977}.
\newblock


\bibitem[\protect\citeauthoryear{Liang and Qian}{Liang and Qian}{2021}]%
        {liang2021recommending}
\bibfield{author}{\bibinfo{person}{Yile Liang} {and} \bibinfo{person}{Tieyun
  Qian}.} \bibinfo{year}{2021}\natexlab{}.
\newblock \showarticletitle{Recommending Accurate and Diverse Items Using
  Bilateral Branch Network}.
\newblock \bibinfo{journal}{\emph{arXiv preprint arXiv:2101.00781}}
  (\bibinfo{year}{2021}).
\newblock


\bibitem[\protect\citeauthoryear{Liu, Chen, Li, Yu, McAuley, and Xiong}{Liu
  et~al\mbox{.}}{2021}]%
        {liu2021contrastive}
\bibfield{author}{\bibinfo{person}{Zhiwei Liu}, \bibinfo{person}{Yongjun Chen},
  \bibinfo{person}{Jia Li}, \bibinfo{person}{Philip~S Yu},
  \bibinfo{person}{Julian McAuley}, {and} \bibinfo{person}{Caiming Xiong}.}
  \bibinfo{year}{2021}\natexlab{}.
\newblock \showarticletitle{Contrastive self-supervised sequential
  recommendation with robust augmentation}.
\newblock \bibinfo{journal}{\emph{arXiv preprint arXiv:2108.06479}}
  (\bibinfo{year}{2021}).
\newblock


\bibitem[\protect\citeauthoryear{McAuley, Targett, Shi, and Van
  Den~Hengel}{McAuley et~al\mbox{.}}{2015}]%
        {mcauley2015image}
\bibfield{author}{\bibinfo{person}{Julian McAuley},
  \bibinfo{person}{Christopher Targett}, \bibinfo{person}{Qinfeng Shi}, {and}
  \bibinfo{person}{Anton Van Den~Hengel}.} \bibinfo{year}{2015}\natexlab{}.
\newblock \showarticletitle{Image-based recommendations on styles and
  substitutes}. In \bibinfo{booktitle}{\emph{Proceedings of the 38th
  international ACM SIGIR conference on research and development in information
  retrieval}}. \bibinfo{pages}{43--52}.
\newblock


\bibitem[\protect\citeauthoryear{Puthiya~Parambath, Usunier, and
  Grandvalet}{Puthiya~Parambath et~al\mbox{.}}{2016}]%
        {puthiya2016coverage}
\bibfield{author}{\bibinfo{person}{Shameem~A Puthiya~Parambath},
  \bibinfo{person}{Nicolas Usunier}, {and} \bibinfo{person}{Yves Grandvalet}.}
  \bibinfo{year}{2016}\natexlab{}.
\newblock \showarticletitle{A coverage-based approach to recommendation
  diversity on similarity graph}. In \bibinfo{booktitle}{\emph{Proceedings of
  the 10th ACM Conference on Recommender Systems}}. \bibinfo{pages}{15--22}.
\newblock


\bibitem[\protect\citeauthoryear{Rendle, Freudenthaler, Gantner, and
  Schmidt-Thieme}{Rendle et~al\mbox{.}}{2012}]%
        {rendle2012bpr}
\bibfield{author}{\bibinfo{person}{Steffen Rendle}, \bibinfo{person}{Christoph
  Freudenthaler}, \bibinfo{person}{Zeno Gantner}, {and} \bibinfo{person}{Lars
  Schmidt-Thieme}.} \bibinfo{year}{2012}\natexlab{}.
\newblock \showarticletitle{BPR: Bayesian personalized ranking from implicit
  feedback}.
\newblock \bibinfo{journal}{\emph{arXiv preprint arXiv:1205.2618}}
  (\bibinfo{year}{2012}).
\newblock


\bibitem[\protect\citeauthoryear{Rendle, Freudenthaler, and
  Schmidt-Thieme}{Rendle et~al\mbox{.}}{2010}]%
        {rendle2010factorizing}
\bibfield{author}{\bibinfo{person}{Steffen Rendle}, \bibinfo{person}{Christoph
  Freudenthaler}, {and} \bibinfo{person}{Lars Schmidt-Thieme}.}
  \bibinfo{year}{2010}\natexlab{}.
\newblock \showarticletitle{Factorizing personalized markov chains for
  next-basket recommendation}. In \bibinfo{booktitle}{\emph{Proceedings of the
  19th international conference on World wide web}}. \bibinfo{pages}{811--820}.
\newblock


\bibitem[\protect\citeauthoryear{Ruby and Yendapalli}{Ruby and
  Yendapalli}{2020}]%
        {ruby2020binary}
\bibfield{author}{\bibinfo{person}{U Ruby} {and} \bibinfo{person}{V
  Yendapalli}.} \bibinfo{year}{2020}\natexlab{}.
\newblock \showarticletitle{Binary cross entropy with deep learning technique
  for image classification}.
\newblock \bibinfo{journal}{\emph{International Journal of Advanced Trends in
  Computer Science and Engineering}} \bibinfo{volume}{9}, \bibinfo{number}{10}
  (\bibinfo{year}{2020}).
\newblock


\bibitem[\protect\citeauthoryear{Socher, Chen, Manning, and Ng}{Socher
  et~al\mbox{.}}{2013}]%
        {socher2013reasoning}
\bibfield{author}{\bibinfo{person}{Richard Socher}, \bibinfo{person}{Danqi
  Chen}, \bibinfo{person}{Christopher~D Manning}, {and} \bibinfo{person}{Andrew
  Ng}.} \bibinfo{year}{2013}\natexlab{}.
\newblock \showarticletitle{Reasoning with neural tensor networks for knowledge
  base completion}.
\newblock \bibinfo{journal}{\emph{Advances in neural information processing
  systems}}  \bibinfo{volume}{26} (\bibinfo{year}{2013}).
\newblock


\bibitem[\protect\citeauthoryear{Song, Yan, Feng, Zhang, Zhao, and Zhang}{Song
  et~al\mbox{.}}{2018}]%
        {song2018towards}
\bibfield{author}{\bibinfo{person}{Yiping Song}, \bibinfo{person}{Rui Yan},
  \bibinfo{person}{Yansong Feng}, \bibinfo{person}{Yaoyuan Zhang},
  \bibinfo{person}{Dongyan Zhao}, {and} \bibinfo{person}{Ming Zhang}.}
  \bibinfo{year}{2018}\natexlab{}.
\newblock \showarticletitle{Towards a neural conversation model with diversity
  net using determinantal point processes}. In
  \bibinfo{booktitle}{\emph{Proceedings of the AAAI Conference on Artificial
  Intelligence}}, Vol.~\bibinfo{volume}{32}.
\newblock


\bibitem[\protect\citeauthoryear{Steck}{Steck}{2015}]%
        {steck2015gaussian}
\bibfield{author}{\bibinfo{person}{Harald Steck}.}
  \bibinfo{year}{2015}\natexlab{}.
\newblock \showarticletitle{Gaussian ranking by matrix factorization}. In
  \bibinfo{booktitle}{\emph{Proceedings of the 9th ACM Conference on
  Recommender Systems}}. \bibinfo{pages}{115--122}.
\newblock


\bibitem[\protect\citeauthoryear{Sun, Liu, Wu, Pei, Lin, Ou, and Jiang}{Sun
  et~al\mbox{.}}{2019}]%
        {sun2019bert4rec}
\bibfield{author}{\bibinfo{person}{Fei Sun}, \bibinfo{person}{Jun Liu},
  \bibinfo{person}{Jian Wu}, \bibinfo{person}{Changhua Pei},
  \bibinfo{person}{Xiao Lin}, \bibinfo{person}{Wenwu Ou}, {and}
  \bibinfo{person}{Peng Jiang}.} \bibinfo{year}{2019}\natexlab{}.
\newblock \showarticletitle{BERT4Rec: Sequential recommendation with
  bidirectional encoder representations from transformer}. In
  \bibinfo{booktitle}{\emph{Proceedings of the 28th ACM international
  conference on information and knowledge management}}.
  \bibinfo{pages}{1441--1450}.
\newblock


\bibitem[\protect\citeauthoryear{Tang and Wang}{Tang and Wang}{2018}]%
        {tang2018personalized}
\bibfield{author}{\bibinfo{person}{Jiaxi Tang} {and} \bibinfo{person}{Ke
  Wang}.} \bibinfo{year}{2018}\natexlab{}.
\newblock \showarticletitle{Personalized top-n sequential recommendation via
  convolutional sequence embedding}. In \bibinfo{booktitle}{\emph{Proceedings
  of the Eleventh ACM International Conference on Web Search and Data Mining}}.
  \bibinfo{pages}{565--573}.
\newblock


\bibitem[\protect\citeauthoryear{Wang, Zhu, Zhu, Qin, and Xiong}{Wang
  et~al\mbox{.}}{2020}]%
        {wang2020setrank}
\bibfield{author}{\bibinfo{person}{Chao Wang}, \bibinfo{person}{Hengshu Zhu},
  \bibinfo{person}{Chen Zhu}, \bibinfo{person}{Chuan Qin}, {and}
  \bibinfo{person}{Hui Xiong}.} \bibinfo{year}{2020}\natexlab{}.
\newblock \showarticletitle{SetRank: A setwise Bayesian approach for
  collaborative ranking from implicit feedback}. In
  \bibinfo{booktitle}{\emph{Proceedings of the AAAI Conference on Artificial
  Intelligence}}, Vol.~\bibinfo{volume}{34}. \bibinfo{pages}{6127--6136}.
\newblock


\bibitem[\protect\citeauthoryear{Wang, Huang, Xu, Shen, and Nie}{Wang
  et~al\mbox{.}}{2018}]%
        {wang2018chat}
\bibfield{author}{\bibinfo{person}{Wenjie Wang}, \bibinfo{person}{Minlie
  Huang}, \bibinfo{person}{Xin-Shun Xu}, \bibinfo{person}{Fumin Shen}, {and}
  \bibinfo{person}{Liqiang Nie}.} \bibinfo{year}{2018}\natexlab{}.
\newblock \showarticletitle{Chat more: Deepening and widening the chatting
  topic via a deep model}. In \bibinfo{booktitle}{\emph{The 41st international
  acm sigir conference on research \& development in information retrieval}}.
  \bibinfo{pages}{255--264}.
\newblock


\bibitem[\protect\citeauthoryear{Wang, He, Wang, Feng, and Chua}{Wang
  et~al\mbox{.}}{2019}]%
        {wang2019neural}
\bibfield{author}{\bibinfo{person}{Xiang Wang}, \bibinfo{person}{Xiangnan He},
  \bibinfo{person}{Meng Wang}, \bibinfo{person}{Fuli Feng}, {and}
  \bibinfo{person}{Tat-Seng Chua}.} \bibinfo{year}{2019}\natexlab{}.
\newblock \showarticletitle{Neural graph collaborative filtering}. In
  \bibinfo{booktitle}{\emph{Proceedings of the 42nd international ACM SIGIR
  conference on Research and development in Information Retrieval}}.
  \bibinfo{pages}{165--174}.
\newblock


\bibitem[\protect\citeauthoryear{Warlop, Mary, and Gartrell}{Warlop
  et~al\mbox{.}}{2019}]%
        {warlop2019tensorized}
\bibfield{author}{\bibinfo{person}{Romain Warlop},
  \bibinfo{person}{J{\'e}r{\'e}mie Mary}, {and} \bibinfo{person}{Mike
  Gartrell}.} \bibinfo{year}{2019}\natexlab{}.
\newblock \showarticletitle{Tensorized determinantal point processes for
  recommendation}. In \bibinfo{booktitle}{\emph{Proceedings of the 25th ACM
  SIGKDD International Conference on Knowledge Discovery \& Data Mining}}.
  \bibinfo{pages}{1605--1615}.
\newblock


\bibitem[\protect\citeauthoryear{Wilhelm, Ramanathan, Bonomo, Jain, Chi, and
  Gillenwater}{Wilhelm et~al\mbox{.}}{2018}]%
        {wilhelm2018practical}
\bibfield{author}{\bibinfo{person}{Mark Wilhelm}, \bibinfo{person}{Ajith
  Ramanathan}, \bibinfo{person}{Alexander Bonomo}, \bibinfo{person}{Sagar
  Jain}, \bibinfo{person}{Ed~H Chi}, {and} \bibinfo{person}{Jennifer
  Gillenwater}.} \bibinfo{year}{2018}\natexlab{}.
\newblock \showarticletitle{Practical diversified recommendations on youtube
  with determinantal point processes}. In \bibinfo{booktitle}{\emph{Proceedings
  of the 27th ACM International Conference on Information and Knowledge
  Management}}. \bibinfo{pages}{2165--2173}.
\newblock


\bibitem[\protect\citeauthoryear{Wu, Gao, Gao, Weng, and Chen}{Wu
  et~al\mbox{.}}{2019a}]%
        {wu2019dual}
\bibfield{author}{\bibinfo{person}{Qitian Wu}, \bibinfo{person}{Yirui Gao},
  \bibinfo{person}{Xiaofeng Gao}, \bibinfo{person}{Paul Weng}, {and}
  \bibinfo{person}{Guihai Chen}.} \bibinfo{year}{2019}\natexlab{a}.
\newblock \showarticletitle{Dual sequential prediction models linking
  sequential recommendation and information dissemination}. In
  \bibinfo{booktitle}{\emph{Proceedings of the 25th ACM SIGKDD International
  Conference on Knowledge Discovery \& Data Mining}}.
  \bibinfo{pages}{447--457}.
\newblock


\bibitem[\protect\citeauthoryear{Wu, Liu, Miao, Zhao, Zhao, and Guan}{Wu
  et~al\mbox{.}}{2019b}]%
        {wu2019pd}
\bibfield{author}{\bibinfo{person}{Qiong Wu}, \bibinfo{person}{Yong Liu},
  \bibinfo{person}{Chunyan Miao}, \bibinfo{person}{Binqiang Zhao},
  \bibinfo{person}{Yin Zhao}, {and} \bibinfo{person}{Lu Guan}.}
  \bibinfo{year}{2019}\natexlab{b}.
\newblock \showarticletitle{PD-GAN: Adversarial Learning for Personalized
  Diversity-Promoting Recommendation.}. In \bibinfo{booktitle}{\emph{IJCAI}},
  Vol.~\bibinfo{volume}{19}. \bibinfo{pages}{3870--3876}.
\newblock


\bibitem[\protect\citeauthoryear{Xie, Sun, Liu, Wu, Gao, Ding, and Cui}{Xie
  et~al\mbox{.}}{2020}]%
        {xie2020contrastive}
\bibfield{author}{\bibinfo{person}{Xu Xie}, \bibinfo{person}{Fei Sun},
  \bibinfo{person}{Zhaoyang Liu}, \bibinfo{person}{Shiwen Wu},
  \bibinfo{person}{Jinyang Gao}, \bibinfo{person}{Bolin Ding}, {and}
  \bibinfo{person}{Bin Cui}.} \bibinfo{year}{2020}\natexlab{}.
\newblock \showarticletitle{Contrastive Learning for Sequential
  Recommendation}.
\newblock \bibinfo{journal}{\emph{arXiv preprint arXiv:2010.14395}}
  (\bibinfo{year}{2020}).
\newblock


\bibitem[\protect\citeauthoryear{Xie, Liu, Zhang, Lu, Wang, and Ding}{Xie
  et~al\mbox{.}}{2021}]%
        {xie2021adversarial}
\bibfield{author}{\bibinfo{person}{Zhe Xie}, \bibinfo{person}{Chengxuan Liu},
  \bibinfo{person}{Yichi Zhang}, \bibinfo{person}{Hongtao Lu},
  \bibinfo{person}{Dong Wang}, {and} \bibinfo{person}{Yue Ding}.}
  \bibinfo{year}{2021}\natexlab{}.
\newblock \showarticletitle{Adversarial and Contrastive Variational Autoencoder
  for Sequential Recommendation}. In \bibinfo{booktitle}{\emph{Proceedings of
  the Web Conference 2021}}. \bibinfo{pages}{449--459}.
\newblock


\bibitem[\protect\citeauthoryear{Xu, Zhao, Liu, Xu, S.~Sheng, Cui, Zhou, and
  Xiong}{Xu et~al\mbox{.}}{2019}]%
        {xu2019recurrent}
\bibfield{author}{\bibinfo{person}{Chengfeng Xu}, \bibinfo{person}{Pengpeng
  Zhao}, \bibinfo{person}{Yanchi Liu}, \bibinfo{person}{Jiajie Xu},
  \bibinfo{person}{Victor S~Sheng S.~Sheng}, \bibinfo{person}{Zhiming Cui},
  \bibinfo{person}{Xiaofang Zhou}, {and} \bibinfo{person}{Hui Xiong}.}
  \bibinfo{year}{2019}\natexlab{}.
\newblock \showarticletitle{Recurrent convolutional neural network for
  sequential recommendation}. In \bibinfo{booktitle}{\emph{The world wide web
  conference}}. \bibinfo{pages}{3398--3404}.
\newblock


\bibitem[\protect\citeauthoryear{Yang, Yu, Zheng, Yao, and Mei}{Yang
  et~al\mbox{.}}{2019}]%
        {yang2019adaptive}
\bibfield{author}{\bibinfo{person}{Shuo Yang}, \bibinfo{person}{Wei Yu},
  \bibinfo{person}{Ying Zheng}, \bibinfo{person}{Hongxun Yao}, {and}
  \bibinfo{person}{Tao Mei}.} \bibinfo{year}{2019}\natexlab{}.
\newblock \showarticletitle{Adaptive semantic-visual tree for hierarchical
  embeddings}. In \bibinfo{booktitle}{\emph{Proceedings of the 27th ACM
  International Conference on Multimedia}}. \bibinfo{pages}{2097--2105}.
\newblock


\bibitem[\protect\citeauthoryear{Yu, Zhang, Liang, and Zhang}{Yu
  et~al\mbox{.}}{2019}]%
        {yu2019multi}
\bibfield{author}{\bibinfo{person}{Lu Yu}, \bibinfo{person}{Chuxu Zhang},
  \bibinfo{person}{Shangsong Liang}, {and} \bibinfo{person}{Xiangliang Zhang}.}
  \bibinfo{year}{2019}\natexlab{}.
\newblock \showarticletitle{Multi-order attentive ranking model for sequential
  recommendation}. In \bibinfo{booktitle}{\emph{Proceedings of the AAAI
  Conference on Artificial Intelligence}}, Vol.~\bibinfo{volume}{33}.
  \bibinfo{pages}{5709--5716}.
\newblock


\bibitem[\protect\citeauthoryear{Yuan, Karatzoglou, Arapakis, Jose, and
  He}{Yuan et~al\mbox{.}}{2019}]%
        {yuan2019simple}
\bibfield{author}{\bibinfo{person}{Fajie Yuan}, \bibinfo{person}{Alexandros
  Karatzoglou}, \bibinfo{person}{Ioannis Arapakis}, \bibinfo{person}{Joemon~M
  Jose}, {and} \bibinfo{person}{Xiangnan He}.} \bibinfo{year}{2019}\natexlab{}.
\newblock \showarticletitle{A simple convolutional generative network for next
  item recommendation}. In \bibinfo{booktitle}{\emph{Proceedings of the Twelfth
  ACM International Conference on Web Search and Data Mining}}.
  \bibinfo{pages}{582--590}.
\newblock


\end{thebibliography}

%%
%% If your work has an appendix, this is the place to put it.

\end{document}